\documentclass[12pt]{iopart}

\usepackage{graphics}
\begin{document}
\title [Comprehensive Analysis of Neutrinos in SK part I]
{
The Comprehensive Analysis of Neutrino Events Occurring inside the
Detector in the Super-Kamiokande Experiment from the View Point of 
the Numerical Computer Experiments: Part~1\\[20pt]
{\large
   --Mutual Relation between the Directions of the Incident
 Neutrinos and those of the Produced Leptons--
}
}
\author{
E. Konishi$^1$, Y. Minorikawa$^2$, V.I. Galkin$^3$, M. Ishiwata$^4$,\\
I. Nakamura$^4$, N. Takahashi$^1$, M. Kato$^5$ and A. Misaki$^6$
}

\address{
$^1$ Graduate School of Science and Technology, Hirosaki University, Hirosaki, 036-8561, Japan }    
\address{
$^2$ Department of Science, School of Science and Engineering, Kinki University, Higashi-Osaka, 577-8502, Japan }
\address{
$^3$ Department of Physics, Moscow State University, Moscow, 119992, Russia}
\address{
$^4$ Department of Physics, Saitama University, Saitama, 338-8570, Japan}
\address{
$^5$ Kyowa Interface Science Co.,Ltd., Saitama, 351-0033, Japan }
\address{
$^6$ Research Institute for Science and Engineering, Waseda University, Tokyo, 169-0092, Japan }
\ead{konish@si.hirosaki-u.ac.jp}

\begin{abstract} 
Super-Kamiokande collaboration assumes that the direction of every 
observed lepton coincides with the incoming direction of the 
incident neutrino, which is the fundamental basement throughout 
all their analysis on neutrino oscillation. 
We examine whether this assumption to explain the experimental 
results on neutrino oscillation is theoretically acceptable. 
Treating every physical process concerned stochastically, we 
have examined if this assumption just cited is acceptable. 
As the result of it, we have shown that this assumption does not 
hold even if statistically.   
\end{abstract}
\pacs{ 13.15.+g, 14.60.-z}
\noindent{\it Keywords}: 
Super-Kamiokande, QEL, Computer Numerical Experiment


\maketitle

\section{Introduction: 
The motivation of the paper}
According to the results obtained from the Super-Kamiokande Experiments 
on atmospheric neutrinos, oscillation phenomena have been found between 
two neutrinos, $\nu_{\mu}$and $\nu_{\tau}$.
 Published reports on the confirmation to the oscillation between the 
neutrinos, $\nu_{\mu}$and $\nu_{\tau}$, and the history forgoing to 
these experiments will be 
critically reviewed and details are in the following:  

\begin{itemize}
\item[(1)]During 1980's Kamiokande and IMB observed smaller atmospheric 
neutrino flux ratio $\nu_{\mu}/\nu_e$ than the predicated value
\cite{Hirata}.
\item[(2)]Kamiokande found anomaly in the zenith angle distribution \cite{Hatakeyama}
\item[(3)]Super-Kamiokande found $\nu_{\mu}$-$\nu_{\tau}$ oscillation 
\cite{Kajita2} and \\
Soudan2 and MACRO confirm the Super-Kamiokande result
\cite{Mann}.
\item[(4)]K2K, the first accelerator-based long baseline experiment, 
confirmed atmospheric neutrino oscillation\cite{K2K}.
\item[(5)]MINOS's precision measurement gives the consistent results
with Super-Kamiokande ones\cite{MINOS}.
\end{itemize}

  It is well known that Super-Kamiokande Collaboration
examined all possible types of the
neutrino events, such as, say, Sub-GeV e-like, Multi-GeV e-like,
Sub-GeV $\mu$-like, Multi-GeV $\mu$-like, Multi-ring Sub-GeV $\mu$-like,
 Multi-ring 
Multi-GeV $\mu$-like, PC, {\it Upward Stopping Muon Events} and 
{\it Upward Through Going Muon Events},
in other words, all possible interactions by neutrino, such as, 
elastic and quasielastic scattering, single-meson production and
deep scattering are considered. Furthermore, all topologically 
different types of neutrino events lead the unified numerical 
oscillation parameters, say,  
$\Delta m^2 = 2.4\times 10^{-3}\rm{ eV^2}$ and $sin^2 2\theta=1.0$
\cite{Ashie}. 

  Taking into account all factors mentioned above, it is natural
that the majority should believe the finding of the $\mu-\tau$ 
neutrino oscillation by Super-Kamiokande Collaboration.
 
  However, it should be emphasized strongly that 
the Super-Kamiokande Collaboration put the 
fundamental assumption in the analysis of the atmospheric neutrino
oscillation which is never self-evident and should be carefully examined.
 This assumption is that the direction of the incident 
neutrino is the same as that of emitted lepton(s).

In order to avoid any misunderstanding toward 
the SK assumption on the direction we reproduce this assumption 
from the original SK papers and their related papers in italic.\\
Kajita and Totsuka state:
\begin{quote}
"{\it However, the direction of the neutrino must be estimated from the
reconstructed direction of the products of the neutrino interaction.
 In water Cheren-kov detectors, the direction of an observed lepton is
assumed to be the direction of the neutrino. Fig.11 and 
Fig.12 show the estimated correlation angle between 
neutrinos and leptons as a function of lepton momentum.
 At energies below 400~MeV/c, the lepton direction has little 
correlation with the neutrino direction. The correlation angle 
becomes smaller with increasing lepton momentum. Therefore, 
the zenith angle dependence of the flux as a consequence of 
neutrino oscillation is largely washed out below~400 MeV/c lepton
momentum. With increasing momentum, the effect can be seen more 
clearly.}" \cite{kajita1}
\end{quote}
On the other hand, Ishitsuka states in his Ph.D thesis which
 is exclusively devoted into the L/E analysis of the 
atmospheric neutrino from Super-Kamiokande experiment as follows:
\begin{quote}
" {\it 8.4  Reconstruction of $L_\nu$
\vskip 2mm

Flight length of neutrino is determined from the neutrino incident
zenith angle, although the energy and the flavor are also involved.
 First, the direction of neutrino is estimated for each sample by 
a different way. Then, the neutrino flight lenght is 
calclulated from the zenith angle of the reconstructed direction.
\\
\\
 8.4.1 Reconstruction of Neutrino Direction

\vspace{-2mm}

{\flushleft{\underline 
{FC Single-ring Sample}
}
}

\vspace{2mm}

The direction of neutrino for FC single-ring sample is 
simply assumed to be the same as the reconstructed direction of muon.
Zenith angle of neutrino is reconstructed as follows:
\[
\hspace{2cm}\cos\Theta^{rec}_{\nu}=\cos\Theta_{\mu} \hspace{2cm}(8.17) 
\]
,where $\cos\Theta^{rec}_{\nu}$ and $\cos\Theta_{\mu}$ are 
cosine of the reconstructed zenith angle of neutrino and muon,
respectively.}" \cite{ishitsuka}
\end{quote}
Furthermore, Jung, Kajita {\it et al.} state:
\begin{quote}
"{\it At neutrino energies of more than a few hundred MeV, the 
direction of the reconstructed lepton approximately represents 
the direction of the original neutrino. 
Hence, for detectors near direction of the lepton. Any effects, 
such as neutrino oscillations, that are a function of the neutrino
 flight distance will be manifest in the 
lepton zenith angle distributions.}" \cite{Jung}
\end{quote}

 Hereafter, we call the fundamental assumption by 
the Super-Kamiokande Experiment simply as 
{\it the SK assumption on the direction}. 
\\

In section~2, we discuss treatment on the cross section for 
Quasi-Elastic Scattering(QEL) stochastically which play an 
particularly important role in the analysis of  
{\it Fully Contained Events}.
 In section~3, we treat the effect of azimuthal angle of QEL over the 
zenith angle of the neutrino events which could not be treated 
in the Detector Simulation carried by the Super-Kaomiokande 
Collaboration.
In section~4, we treat the zenith angle distributions of the neutrino 
events for the different directions of the incident neutrinos
stochastically, taking into account of the corresponding incident 
neutrino energy spectrum.
In section~5, we discuss the correlation diagram between the zenith 
angles of the incident neutrinos and those of the produced muons,
and show that {\it the SK assumption on the direction} 
 does not hold even if statistically. 
As the result of it, for example, upward neutrino zenith angle 
 distribution in the absence of neutrino oscillation does not 
reproduce the corresponding one of the produced muons.
 In the conclusion, we state that we will give the results of 
re-analysis of the $L/E$ distribution obtained by 
Super-Kamiokande experiment  under the situation that 
{\it the SK assumption on the direction} 
does not hold in the subsequent paper. 

\section{Cross Sections of Quasi Elastic Scattering in the Neutrino 
Reaction and the Scattering Angle of Charged Leptons.}
 In order to examine the validity of {\it the SK assumption on 
the direction},  we consider the following quasielastic scattering(QEL):

   \begin{eqnarray}
         \nu_e + n \longrightarrow p + e^-  \nonumber\\
         \nu_{\mu} + n \longrightarrow p + \mu^- \nonumber\\
         \bar{\nu}_e + p \longrightarrow n + e^+ \\
         \bar{\nu}_{\mu}+ p \longrightarrow n + \mu^+ \nonumber
         ,\label{eqn:1}
   \end{eqnarray}
because the QEL is the most dominant process in 
{\it Fully Contained Events} and {\it Partially Contained Events} which occur in the detector.
Also, we adopt the QEL events among 
{\it Fully Contained Events} as the object for the analysis, 
because they are simple and solid events in which the energies of 
the produced muons (electrons) could be uniquely determined,
 compared with any other type of the neutrino event. 
The differential cross section for QEL is given as follows \cite{r4}.\\
    \begin{eqnarray}
         \frac{ {\rm d}\sigma_{\ell(\bar{\ell})}(E_{\nu(\bar{\nu})}) }{{\rm d}Q^2} = 
         \frac{G_F^2{\rm cos}^2 \theta_C}{8\pi E_{\nu(\bar{\nu})}^2}
         \Biggl\{ A(Q^2) \pm B(Q^2) \biggl[ \frac{s-u}{M^2} \biggr]
         + \nonumber \\ 
C(Q^2) \biggl[ \frac{s-u}{M^2} \biggr]^2 \Biggr\},
         \label{eqn:2}
    \end{eqnarray}

\noindent where
    \begin{eqnarray*}
      A(Q^2) &=& \frac{Q^2}{4}\Biggl[ f^2_1\biggl( \frac{Q^2}{M^2}-4 \biggr)+ f_1f_2 \frac{4Q^2}{M^2} \\
 &&+  f_2^2\biggl( \frac{Q^2}{M^2} -\frac{Q^4}{4M^4} \biggr) + g_1^2\biggl( 4+\frac{Q^2}{M^2} \biggl) \Biggr], \\
      B(Q^2) &=& (f_1+f_2)g_1Q^2, \\
      C(Q^2) &=& \frac{M^2}{4}\biggl( f^2_1+f^2_2\frac{Q^2}{4M^2}+g_1^2 \biggr).
    \end{eqnarray*}

\noindent The signs $+$ and $-$ refer to $\nu_{\mu(e)}$ and $\bar{\nu}_{\mu(e)}$ for charged current (c.c.) interactions, respectively.  The $Q^2$ denotes the four momentum transfer between the incident neutrino and the charged lepton. Details of other symbols are given in \cite{r4}.

The relation among $Q^2$, $E_{\nu(\bar{\nu})}$, the energy of the incident neutrino, $E_{\ell}$, the energy of the emitted charged lepton (muon or electron or their anti-particles) and $\theta_{\rm s}$, the scattering angle of the emitted lepton, is given as
      \begin{equation}
         Q^2 = 2E_{\nu(\bar{\nu})}E_{\ell}(1-{\rm cos}\theta_{\rm s}).
                  \label{eqn:3}
      \end{equation}

\noindent Also, the energy of the emitted lepton is given by
      \begin{equation}
         E_{\ell} = E_{\nu(\bar{\nu})} - \frac{Q^2}{2M}.
\label{eqn:4}
      \end{equation}

Now, let us examine the magnitude of the scattering angle of the emitted 
lepton in a quantitative way, as this plays a decisive role in determining 
the accuracy of the direction of the incident neutrino,
which is directly related to the reliability of the zenith angle 
distribution of both {\it Fully Contained Events} and {\it Partially 
Contained Events} in the Super-Kamiokande Experiment.

By using Eqs. (\ref{eqn:2}) to (\ref{eqn:4}), we obtain the distribution 
function for the scattering angle of the emitted leptons and the related 
quantities by a Monte Carlo method. The procedure for determining the 
scattering angle for a given energy of the incident neutrino is described 
in the Appendix A.  Fig. \ref{fig:1} shows this relation for muon, from 
which we can easily understand that the scattering angle $\theta_{\rm s}$ 
of the emitted lepton ( muon here ) cannot be neglected.  For a 
quantitative examination of the scattering angle, we construct the 
distribution function for ${\theta_{\rm s}}$ of the emitted lepton from 
Eqs. (\ref{eqn:2}) to (\ref{eqn:4}) by using a Monte Carlo method.

Fig. \ref{fig:2} gives the distribution function for $\theta_{\rm s}$ of 
the muon produced in the muon neutrino interaction. It can be seen that 
the muons produced from lower energy neutrinos are scattered over wider 
angles and that a considerable part of them are scattered even in backward 
directions. 
Similar results are obtained for anti-muon neutrinos, electron neutrinos 
and anti-electron neutrinos.
\begin{figure}
\begin{center}
\vspace{-1cm}
\rotatebox{90}{%
\resizebox{0.5\textwidth}{!}{\includegraphics{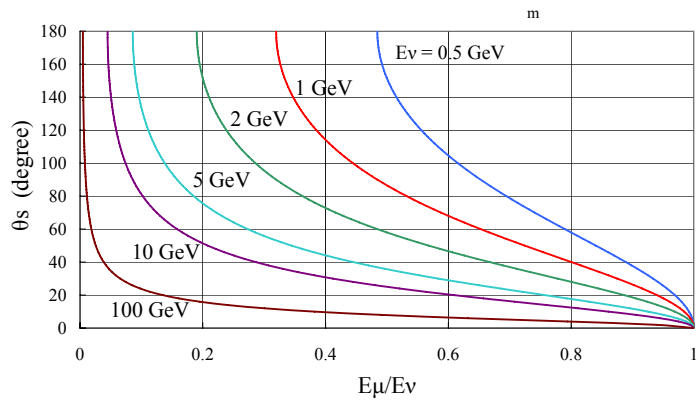}}}
\vspace{-1cm}
\caption{\label{fig:1} Relation between the energy of the muon and its 
scattering angle for different incident muon neutrino energies,
 0.5, 1, 2, 5, 10 and 100~GeV.}
\label{fig:1}
\end{center}
\end{figure} 
\begin{figure}
\begin{center}
\vspace{-1cm}
\rotatebox{90}{%
\resizebox{0.5\textwidth}{!}{\includegraphics{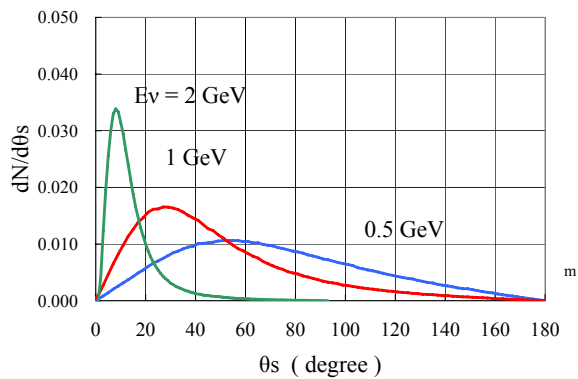}}}
\vspace{-1cm}
\caption{\label{fig:2} Distribution functions for the scattering angle of 
the muon for muon-neutrino with incident energies, 0.5 , 1.0 and 
2~GeV. Each curve is obtained by the Monte Carlo 
method (one million sampling per each curve). }
\label{fig:2}
\end{center}
\end{figure} 

Also, in a similar manner, we obtain not only the distribution function 
for the scattering angle of the charged leptons, but also their average 
values $<\theta_{\rm s}>$ and their standard deviations $\sigma_{\rm s}$. 
Table~1 shows them  for muon neutrinos, anti-muon neutrinos, electron 
neutrinos and anti-electron neutrinos.  
From the Table~1, it seems that the scattering angles could not 
be neglected. However, Super-Kamiokande Collaboration
 approximate them as zero, as cited in three 
passages (in Italic) mentioned above \cite{kajita1},\cite{ishitsuka},\cite{Jung}. 
Or they may claim that such the assumption surely holds statistically
as whole (see the Conclusion of our paper, relating to the latter surmise).

\begin{table*}
\caption{\label{tab:table1} The average values $<\theta_{\rm s}>$ for 
scattering angle of the emitted charged leptons and their standard 
deviations $\sigma_{\rm s}$ for various primary neutrino energies 
$E_{\nu(\bar{\nu})}$.}
\vspace{5mm}
\begin{center}
\begin{tabular}{|c|c|c|c|c|c|}
\hline
&&&&&\\
$E_{\nu(\bar{\nu})}$ (GeV)&angle&$\nu_{\mu(\bar{\mu})}$&
$\bar{\nu}_{\mu(\bar{\mu})}$&$\nu_e$&$\bar{\nu_e}$ \\
&(degree)&&&&\\
\hline
0.2&$<\theta_\mathrm{s}>$&~~ 89.86 ~~&~~ 67.29 ~~&~~ 89.74 ~~&~~ 67.47 ~~\\
\cline{2-6}
   & $\sigma_\mathrm{s}$ & 38.63 & 36.39 & 38.65 & 36.45 \\
\hline
0.5&$<\theta_\mathrm{s}>$& 72.17 & 50.71 & 72.12 & 50.78 \\
\cline{2-6}
   & $\sigma_\mathrm{s}$ & 37.08 & 32.79 & 37.08 & 32.82 \\
\hline
1  &$<\theta_\mathrm{s}>$& 48.44 & 36.00 & 48.42 & 36.01 \\
\cline{2-6}
   & $\sigma_\mathrm{s}$ & 32.07 & 27.05 & 32.06 & 27.05 \\
\hline
2  &$<\theta_\mathrm{s}>$& 25.84 & 20.20 & 25.84 & 20.20 \\
\cline{2-6}
   & $\sigma_\mathrm{s}$ & 21.40 & 17.04 & 21.40 & 17.04 \\
\hline
5  &$<\theta_\mathrm{s}>$&  8.84 &  7.87 &  8.84 &  7.87 \\
\cline{2-6}
   & $\sigma_\mathrm{s}$ &  8.01 &  7.33 &  8.01 &  7.33 \\
\hline
10 &$<\theta_\mathrm{s}>$&  4.14 &  3.82 &  4.14 &  3.82 \\
\cline{2-6}
   & $\sigma_\mathrm{s}$ &  3.71 &  3.22 &  3.71 &  3.22 \\
\hline
100&$<\theta_\mathrm{s}>$&  0.38 &  0.39 &  0.38 &  0.39 \\
\cline{2-6}
   & $\sigma_\mathrm{s}$ &  0.23 &  0.24 &  0.23 &  0.24 \\
\hline
\end{tabular}
\end{center}
\label{tab:1}
\end{table*}

%
\section{Influence of Azimuthal Angle of Quasi Elastic Scattering over the 
Zenith Angle of both Fully Contained Events and 
Partially Contained Events}

In the present section, we examine the effect of the azimuthal angles,
$\phi$, of the emitted leptons over their own zenith angles,
$\theta_{\mu(\bar{\mu}))}$, for given zenith angles 
of the incident neutrinos, $\theta_{\nu(\bar{\nu}))}$,
 which could not be considered in the detector simulation
carried by the Super-Kamiokande Collaboration.
\footnote{Throughout this paper, we measure the zenith angles of the 
emitted leptons from the upward vertical direction of the incident 
neutrino. Consequently, notice that the sign of our direction is oposite 
to that of the Super-Kamiokande Experiment 
( our $\cos\theta_{\nu(\bar{\nu})}$ = - 
$\cos\theta_{\nu(\bar{\nu})}$ in SK)}

For three typical cases (vertical, horizontal and diagonal), Fig. 3 gives 
a schematic representation of the relationship between, 
$\theta_{\nu(\bar{\nu})}$, the zenith angle of the incident neutrino, and (
$\theta_{\rm s}$, $\phi$), a pair of scattering angle of the emitted lepton 
and its azimuthal angle.  

From Fig. 3(a), it can been seen that the zenith angle 
$\theta_{\mu(\bar{\mu})}$ of the emitted lepton is not influenced by its 
$\phi$ in the vertical incidence of the neutrinos 
$(\theta_{\nu(\bar{\nu})}=0^{\rm o})$, as it must be. From Fig. 3(b), 
however, it is obvious that the influence of $\phi$ of the emitted leptons 
on their own zenith angle is the strongest in the case of horizontal 
incidence of the neutrino $(\theta_{\nu(\bar{\nu})}=90^{\rm o})$. Namely, 
one half of the emitted leptons are recognized as upward going, while the 
other half is classified as downward going ones. The diagonal case ( 
$\theta_{\nu(\bar{\nu})}=43^{\rm o}$) is intermediate between the vertical 
and the horizontal. In the following, we examine the cases for vertical, 
horizontal and diagonal incidence of the neutrino with different energies, 
say, $E_{\nu(\bar{\nu})}=0.5$ GeV, $E_{\nu(\bar{\nu})}=1$ GeV and 
$E_{\nu(\bar{\nu})}=5$ GeV,
as the typical cases. 

\begin{figure}
\begin{center}
\vspace{-0.5cm}
\resizebox{0.5\textwidth}{!}{%
  \includegraphics{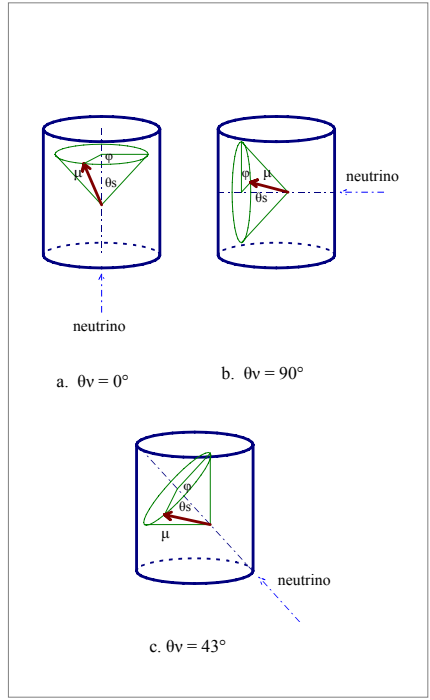}
  }
\end{center}
\caption{\label{fig:3} Schematic view of the zenith angles of the charged
 muons for different zenith angles of the incident neutrinos, focusing on
  their azimuthal angles.}
\label{fig:3}
\end{figure} 

\subsection{Dependence of the spreads of the zenith angle for the emitted 
leptons on the energies 
of the emitted leptons for different incident directions 
of the neutrinos with different 
energies}
\begin{figure*}
\hspace{2cm}(a)
\hspace{5cm}(b)
\hspace{5cm}(c)
\vspace{-0.3cm}
\begin{center}
\resizebox{\textwidth}{!}{%
  \includegraphics{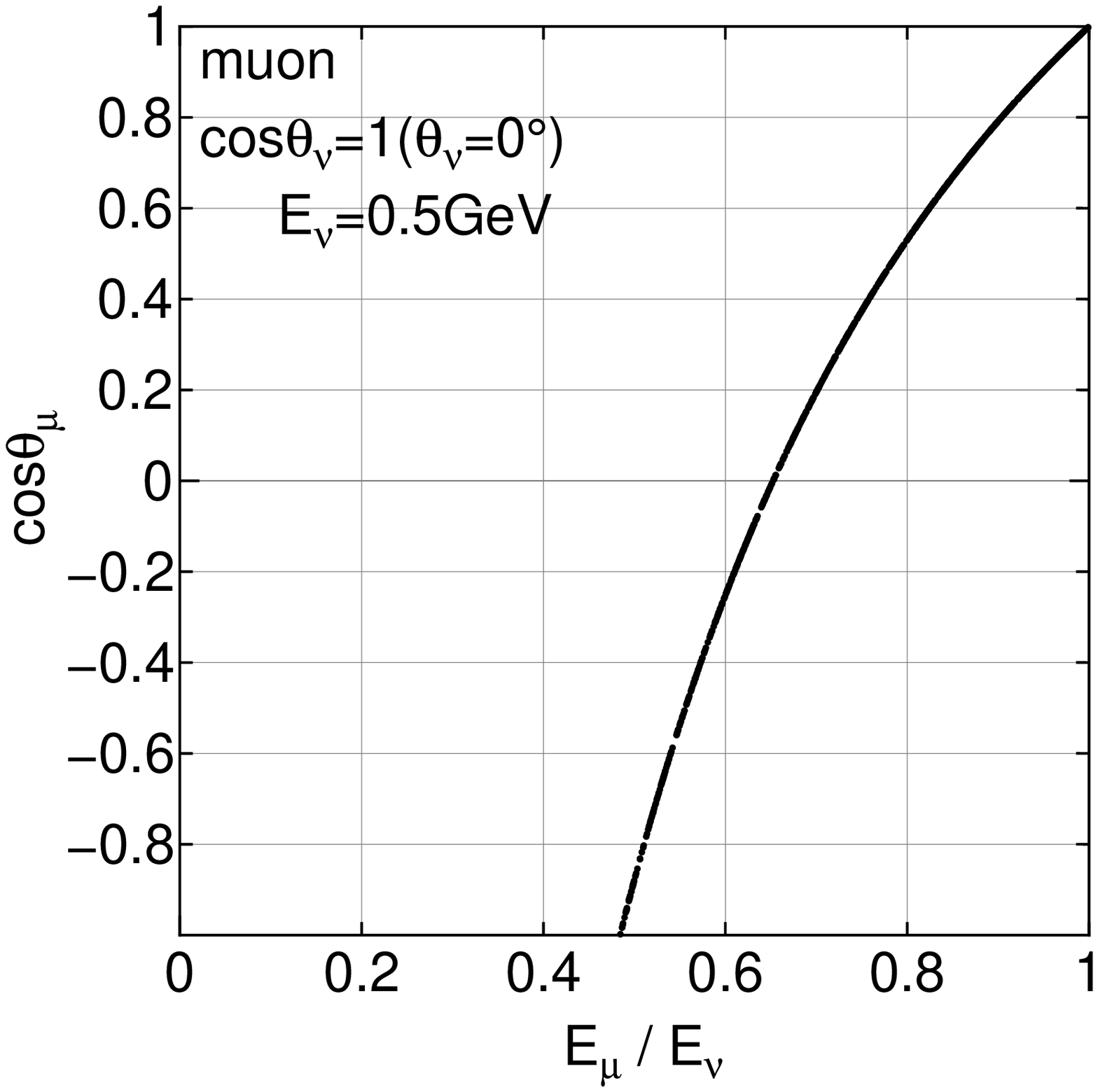}\hspace{1cm}
  \includegraphics{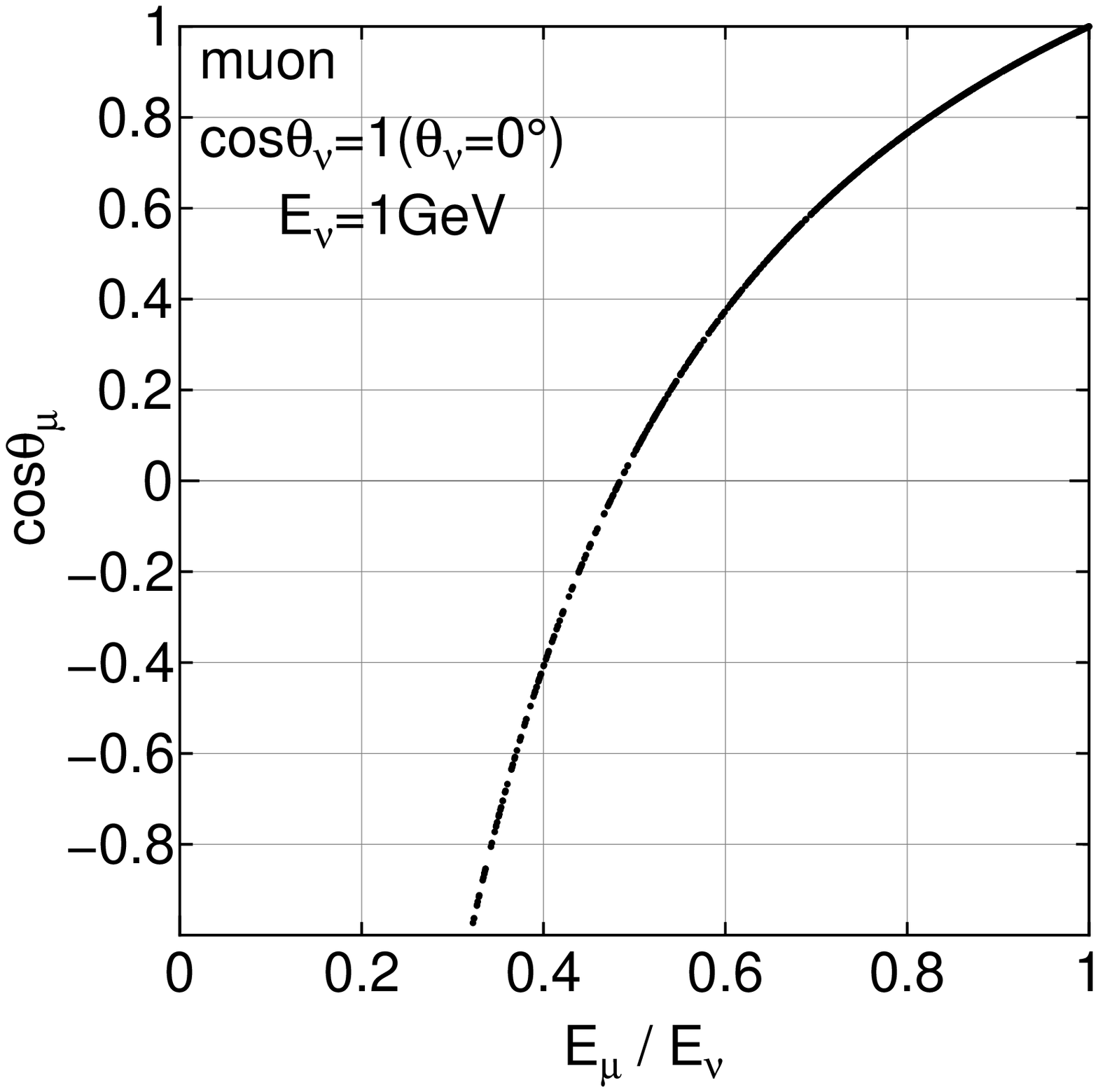}\hspace{1cm}
  \includegraphics{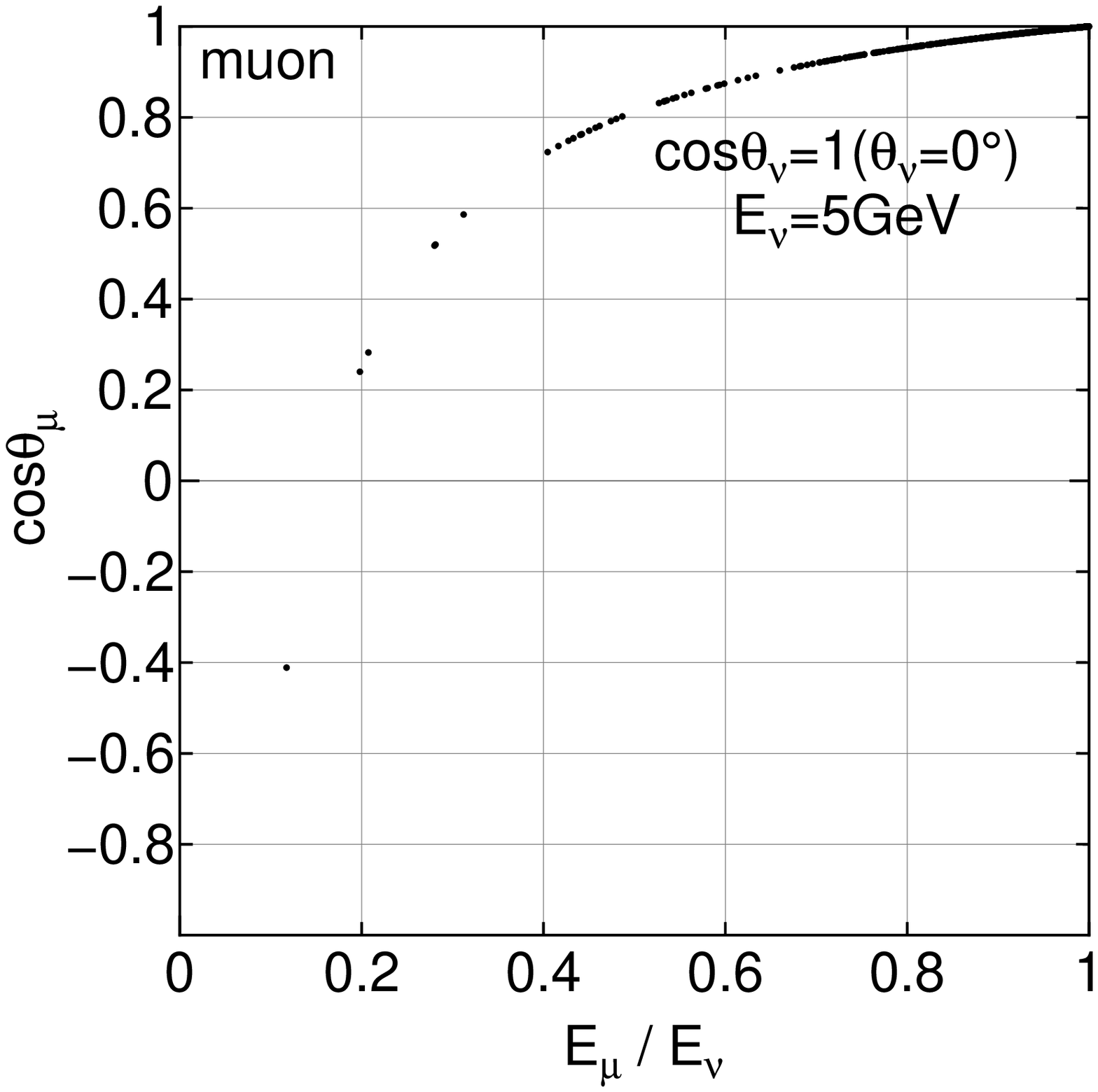}
}
\caption{
\label{fig:4}
The scatter plots between the fractional energies of the produced muons 
and their zenith angles 
for vertically incident muon neutrinos with 0.5~GeV, 1~GeV and 5~GeV, 
respectively.
 The sampling number is 1000 for each case.
}
\end{center}
\vspace{0.5cm}
\hspace{2cm}(a)
\hspace{5cm}(b)
\hspace{5cm}(c)
\vspace{-0.3cm}
\begin{center}
\resizebox{\textwidth}{!}{%
  \includegraphics{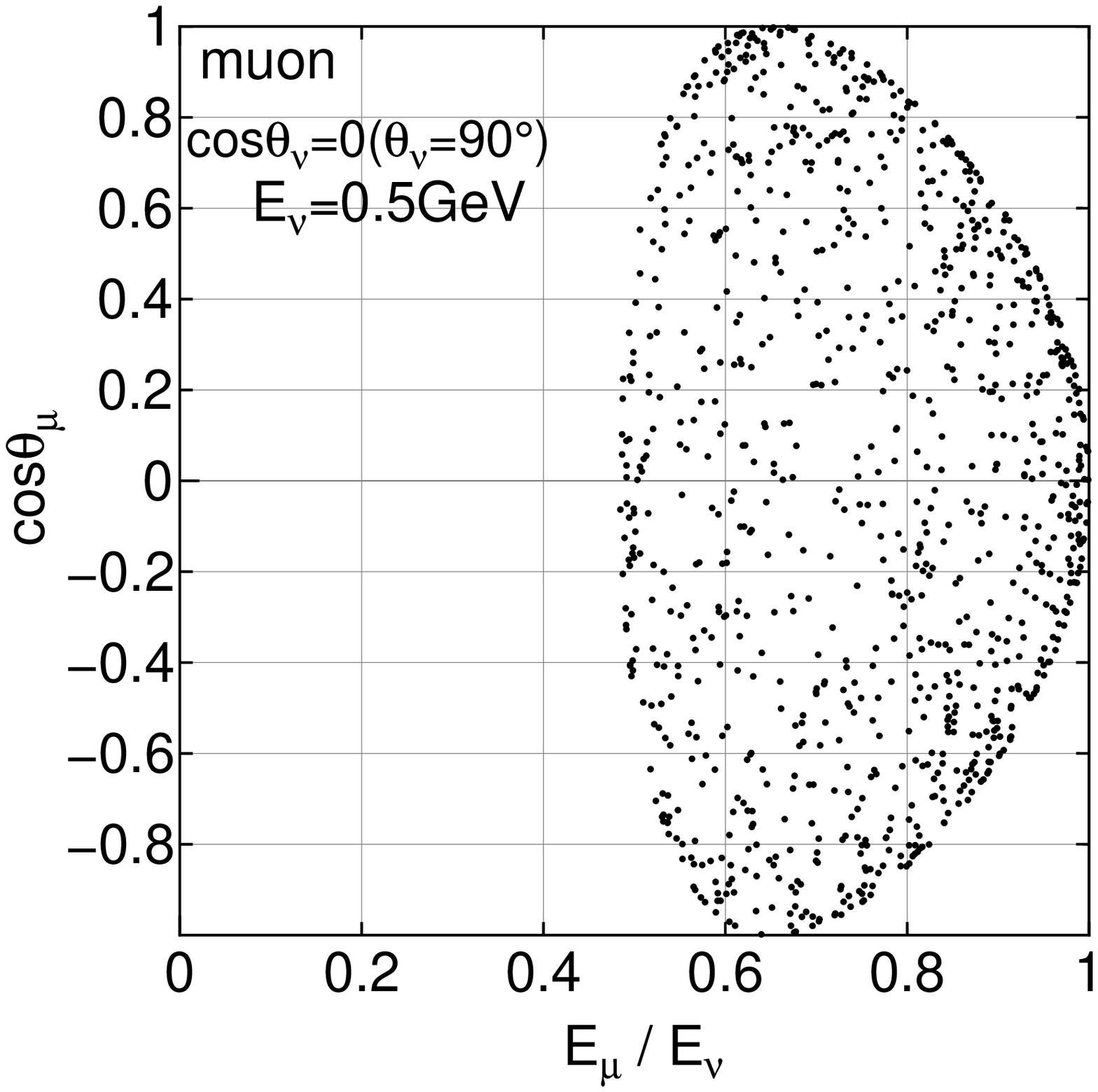}\hspace{1cm}
  \includegraphics{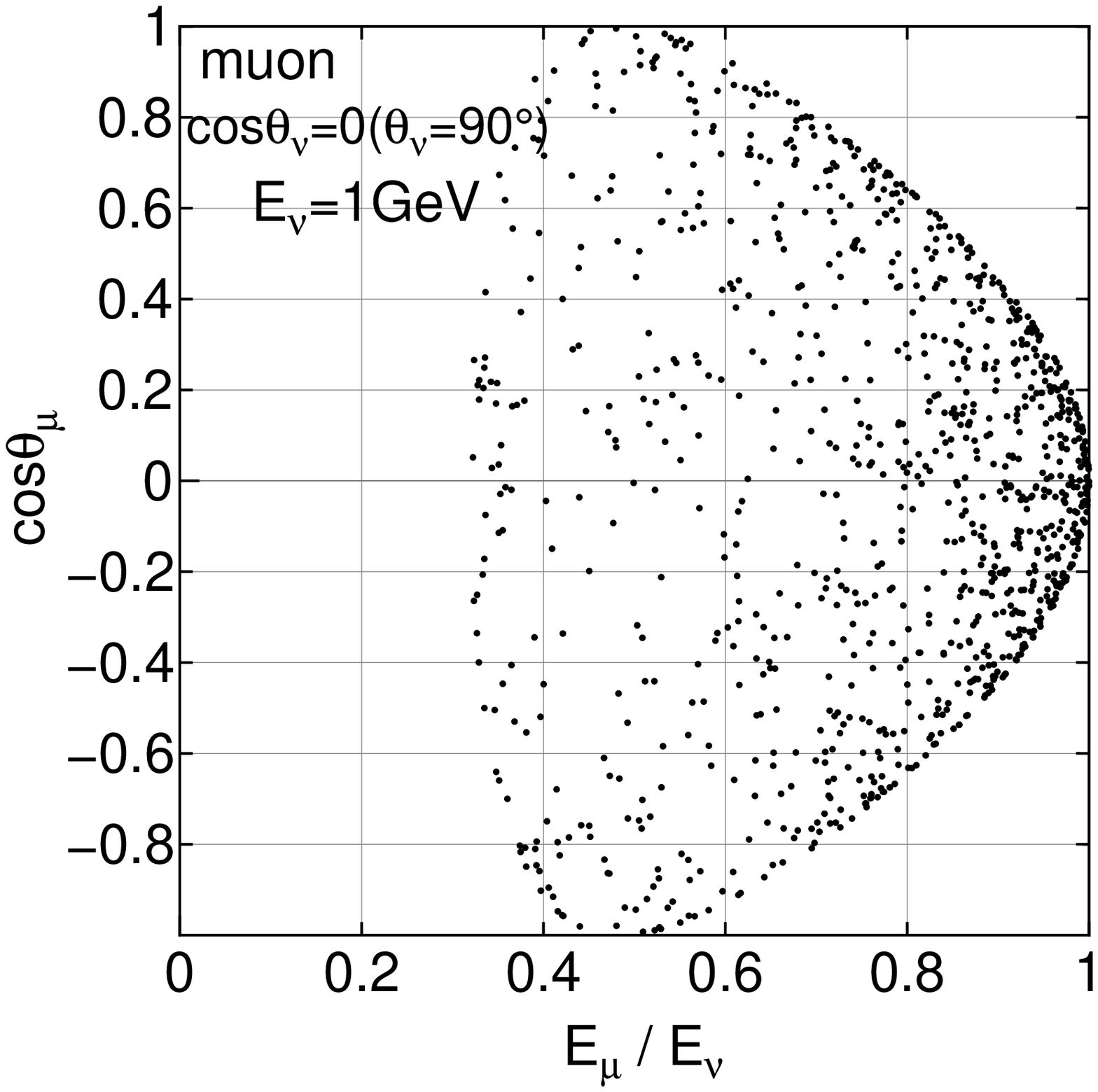}\hspace{1cm}
  \includegraphics{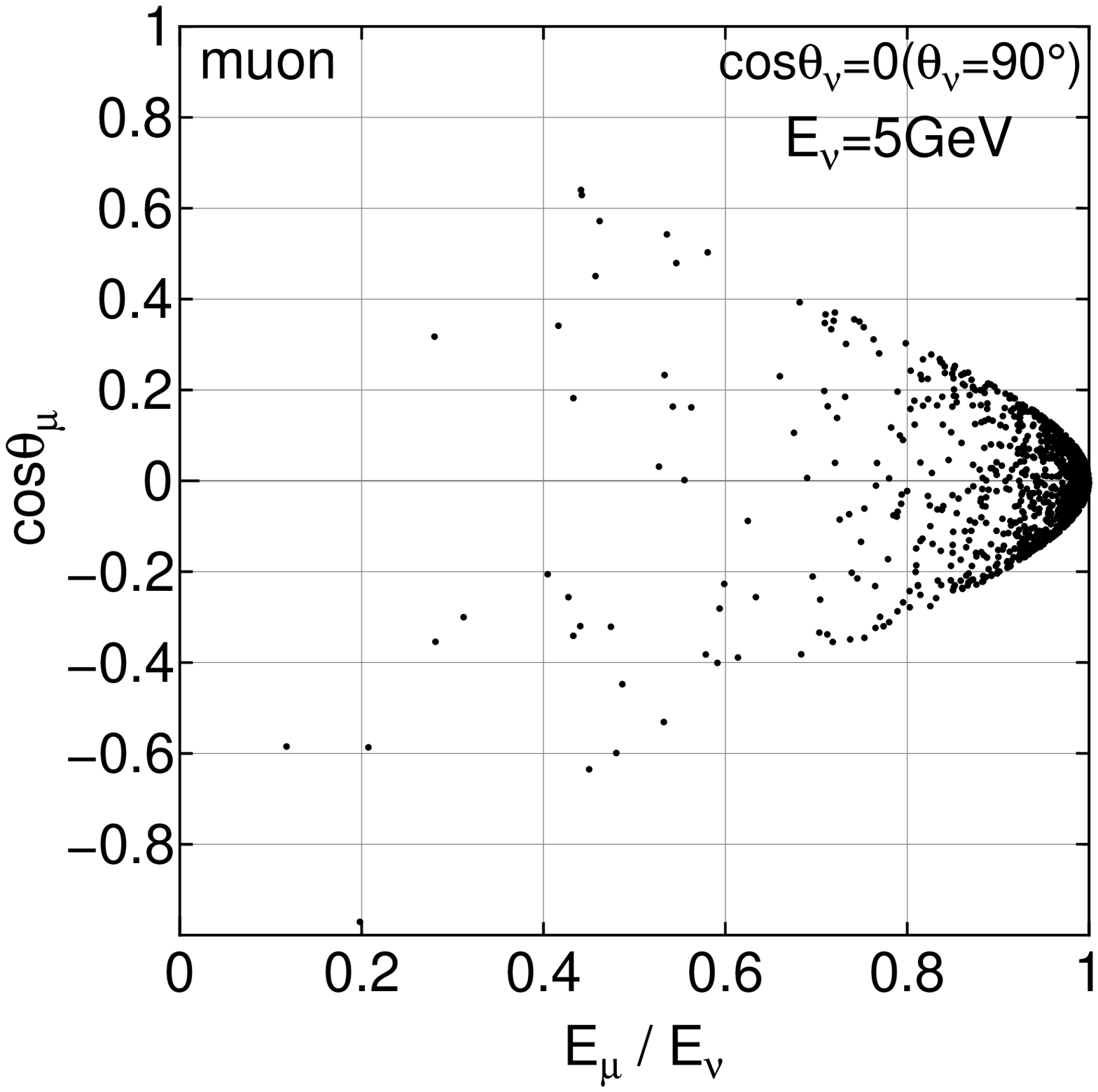}
}
\caption{
\label{fig:5} 
The scatter plots between the fractional energies of the produced muons 
and their zenith angles 
for horizontally incident muon neutrinos with 0.5~GeV, 1~GeV and 5~GeV, 
respectively.
 The sampling number is 1000 for each case.
}
\end{center}
\vspace{0.5cm}
\hspace{2cm}(a)
\hspace{5cm}(b)
\hspace{5cm}(c)
\vspace{-0.3cm}
\begin{center}
\resizebox{\textwidth}{!}{%
  \includegraphics{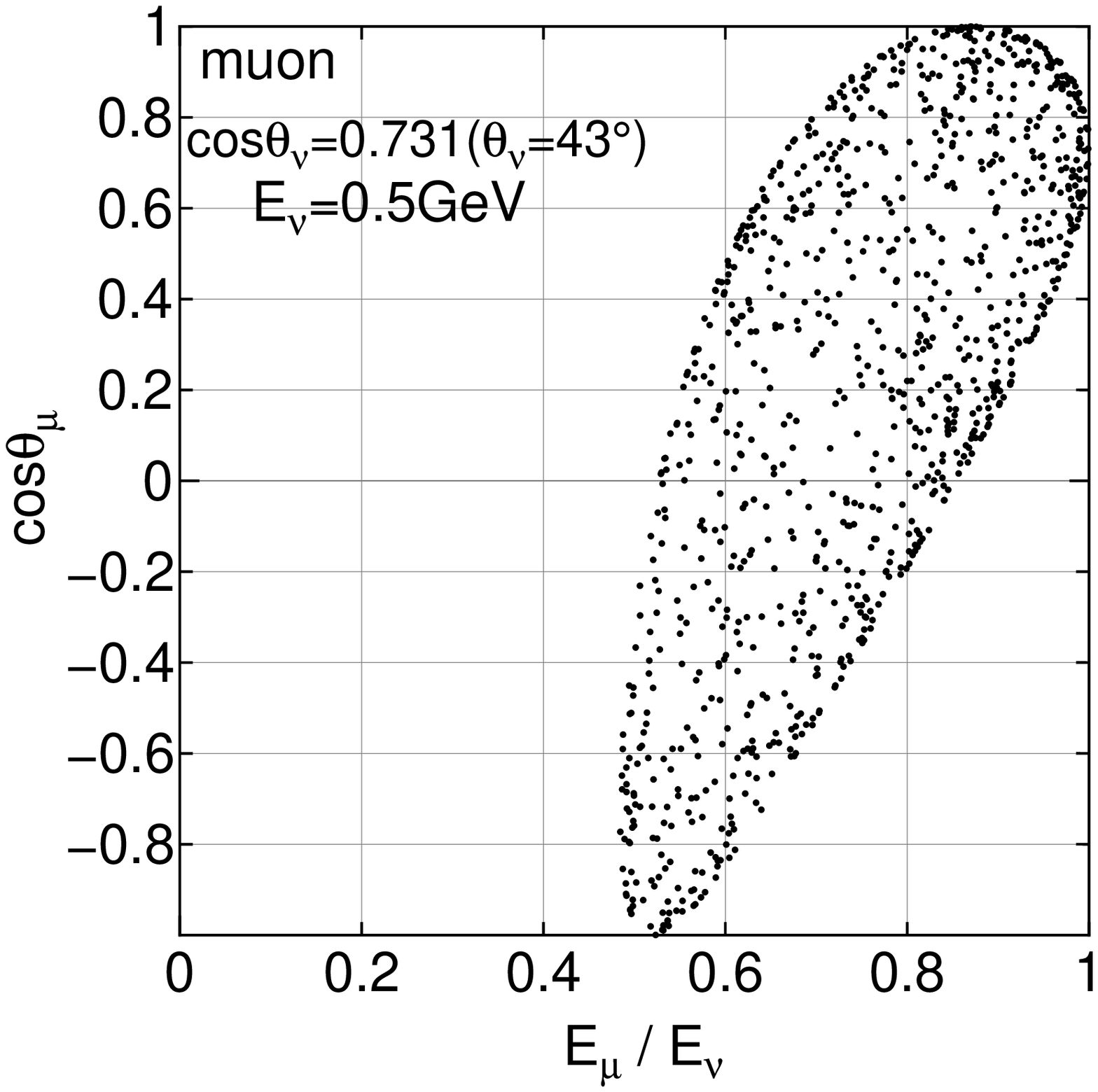}\hspace{1cm}
  \includegraphics{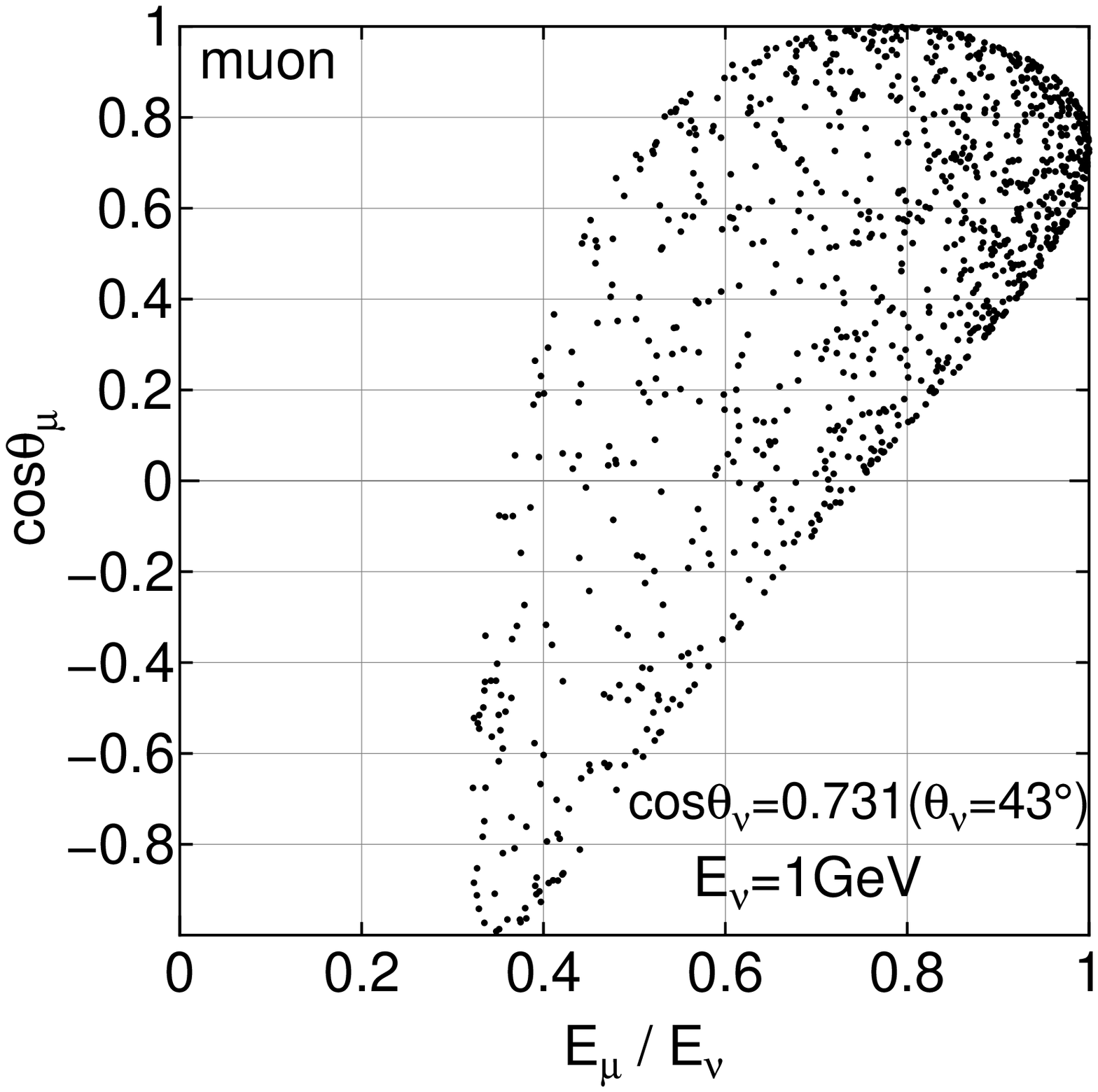}\hspace{1cm}
  \includegraphics{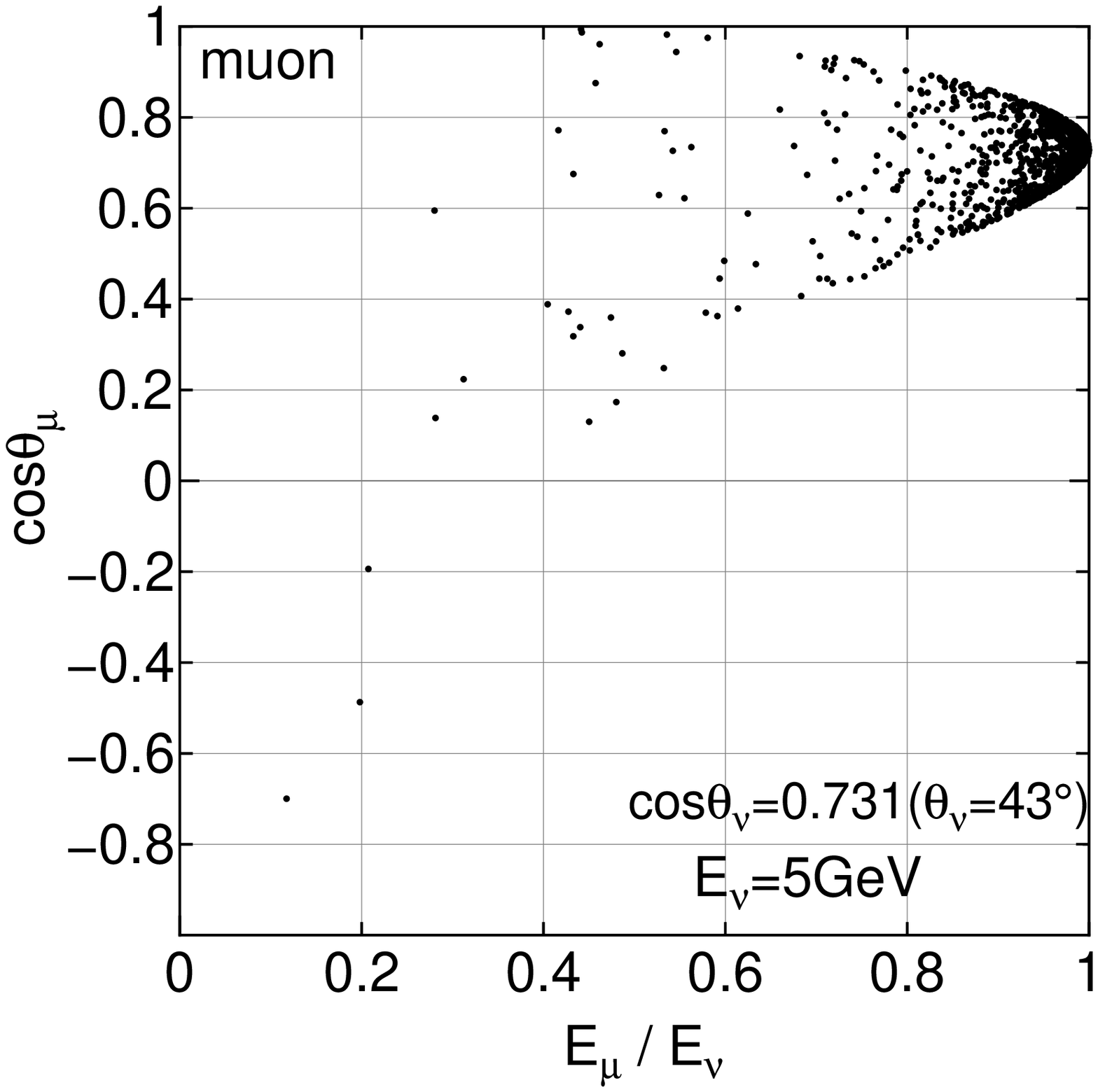}
  }
\caption{
\label{fig:6} 
The scatter plots between the fractional energies of the produced muons 
and their zenith angles 
for diagonally incident muon neutrinos with 0.5~GeV, 1~GeV and 5~GeV, 
respectively.
 The sampling number is 1000 for each case.
}
\end{center}
\end{figure*}

  The detailed procedure for the Monte Carlo simulation is described in 
the Appendix A. 
 We give  the scatter plots between the fractional energies of the emitted 
muons and their zenith angle
 for a definite zenith angles of the incident neutrino with different 
energies in Figs. \ref{fig:4} to \ref{fig:6}. In Fig. \ref{fig:4}, we give 
the scatter plots for vertically incident neutrino with different energies 
0.5 GeV, 1 GeV and 5 GeV . In this case, the relations between the emitted 
energies of the muons and their zenith angles are unique, which comes 
from the definition of the zenith angle of the emitted lepton. However, 
the densities (frequencies of the event number) along each curves are 
different 
in position to position and depend on the energies of the incident 
neutrinos. Generally speaking, densities along the curves become greater 
toward  $\cos\theta_{\mu(\bar{\mu})}= 1$. In this case, 
$\cos\theta_{\mu(\bar{\mu})}$ is never influenced by the azimuthal angel 
in the scattering by the definition
\footnote{The zenith angles of the particles concerned are measured from 
the vertical direction.}.
 

On the contrast, it is shown in Figure~5 that the horizontally incident 
neutrinos give the widest zenith angles distribution for the 
emitted energies of muons due to the effect of their azimuthal angles.
 The more lower incident neutrino energies, the more 
wider spread of the emitted leptons.  As easily understood from Figure~6, 
the diagonally incident neutrinos give the intermediate zenith angle 
distributions for the emitted  muons between those for vertically 
incident neutrinos and those for horizontally neutrinos.  

\begin{figure*}
\hspace{2cm}(a)
\hspace{5cm}(b)
\hspace{5cm}(c)
\vspace{-0.5cm}
\begin{center}
\resizebox{\textwidth}{!}{%
  \includegraphics{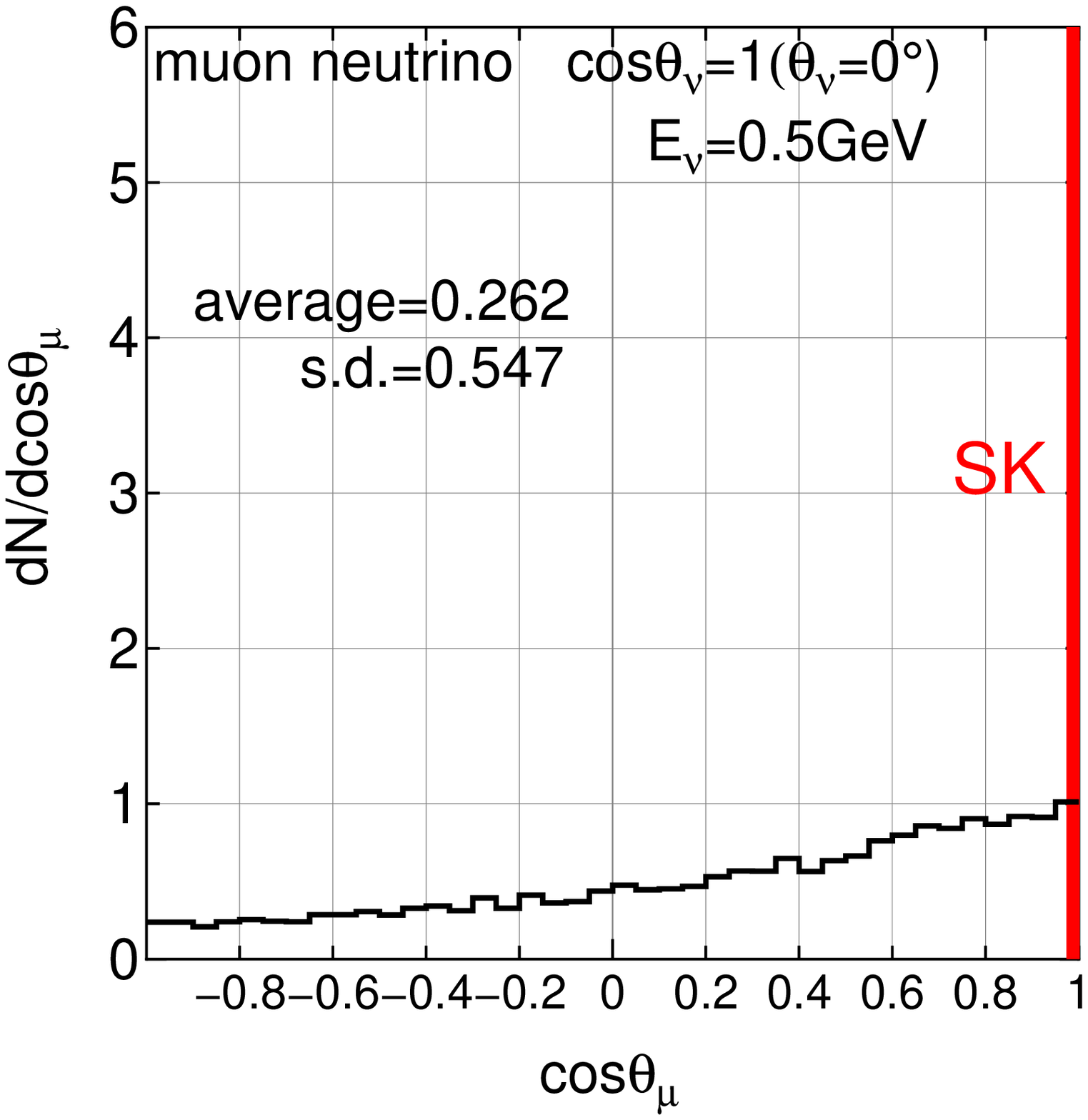}\hspace{1cm}
  \includegraphics{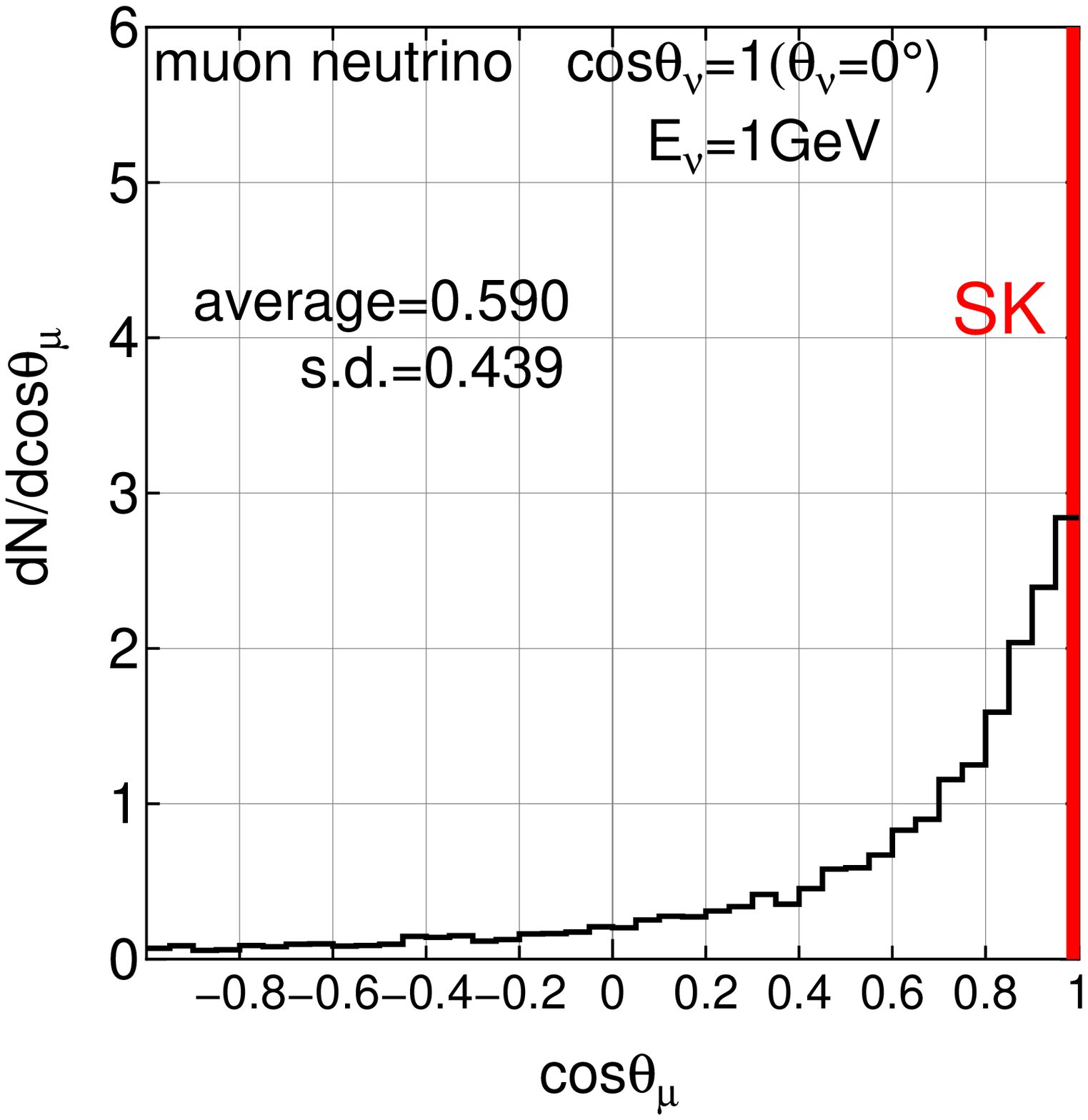}\hspace{1cm}
  \includegraphics{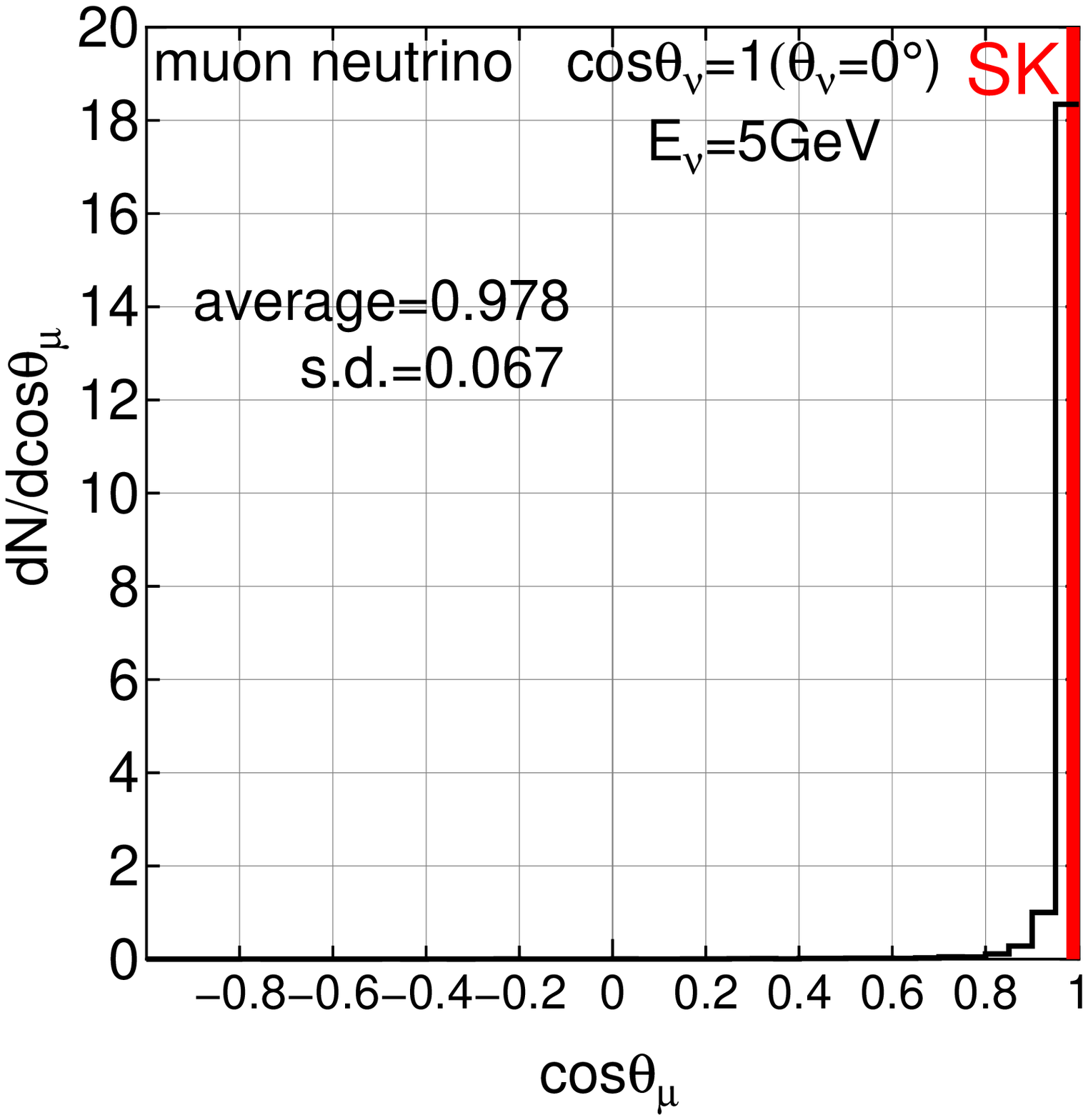}
  }
\caption{
\label{fig:7} 
Zenith angle distribution of the muon for the vertically incident muon 
neutrino with 0.5~GeV, 1~GeV and 5~GeV, respectively. The sampling 
number is 10000 for each case.
SK stands for the corresponding ones under the SK assumption.
}
\end{center}
\vspace{0.5cm}

\hspace{2cm}(a)
\hspace{5cm}(b)
\hspace{5cm}(c)
\vspace{-0.5cm}
\begin{center}
\resizebox{\textwidth}{!}{%
 \includegraphics{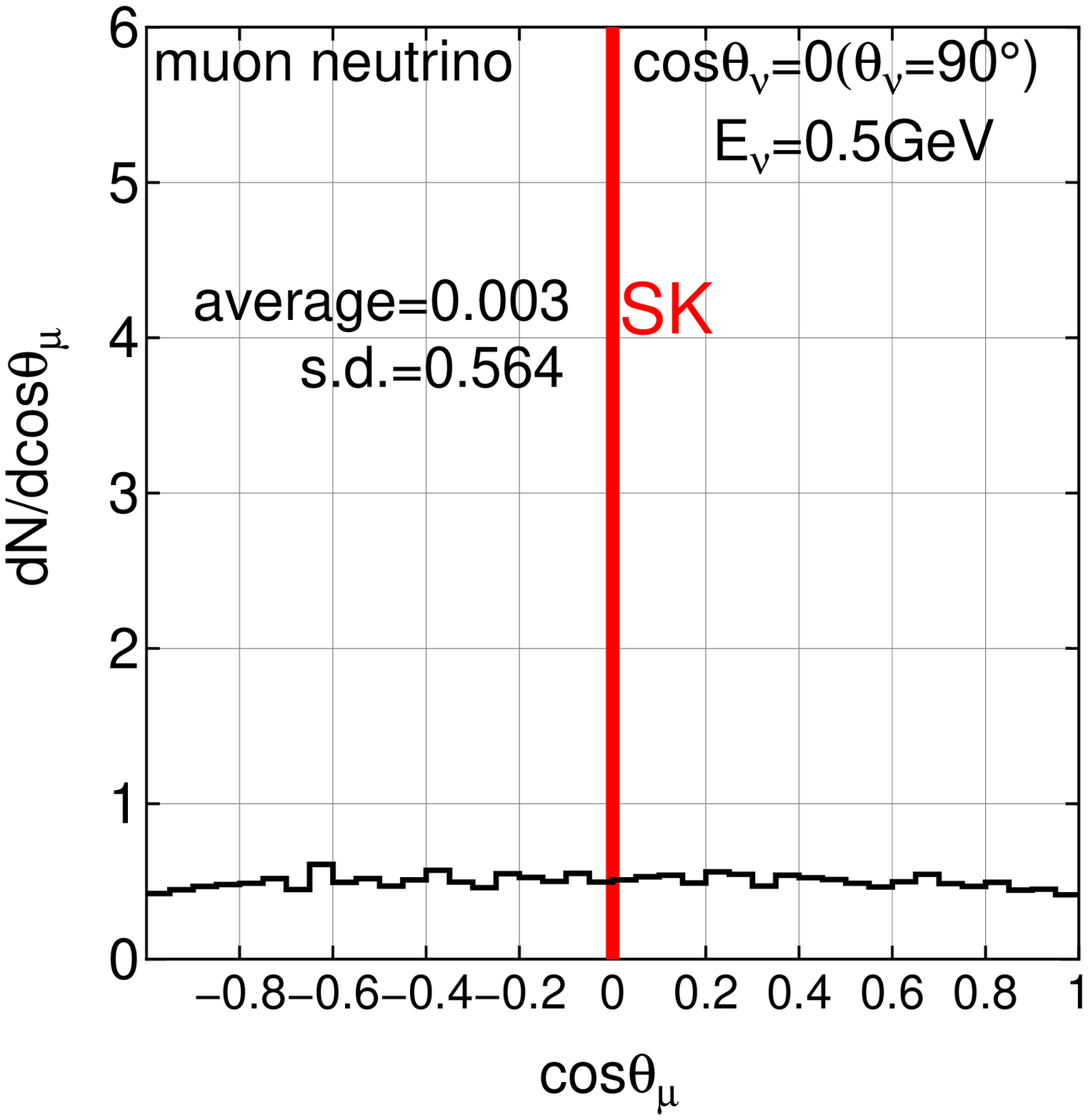}\hspace{1cm}
  \includegraphics{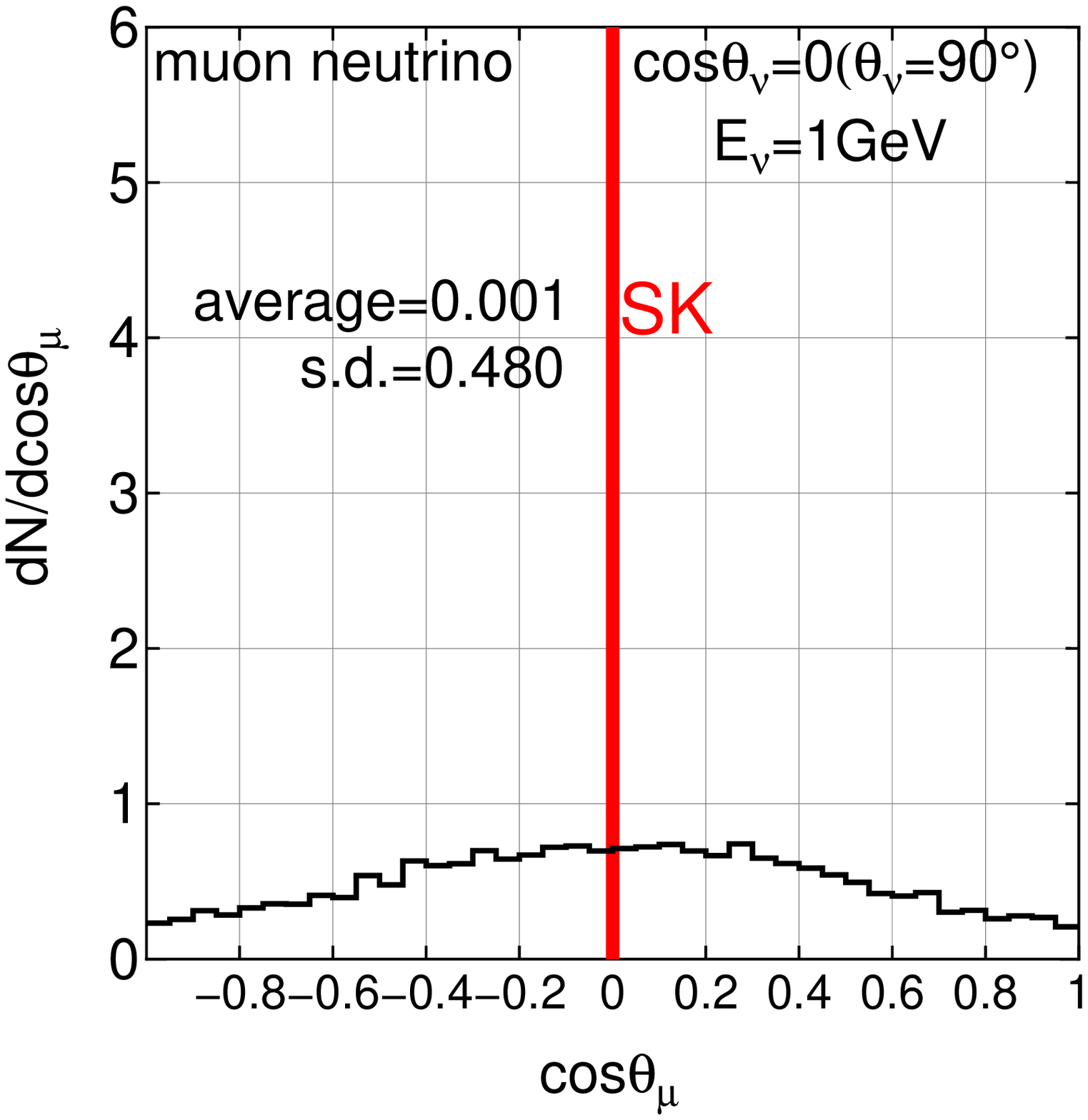}\hspace{1cm}
  \includegraphics{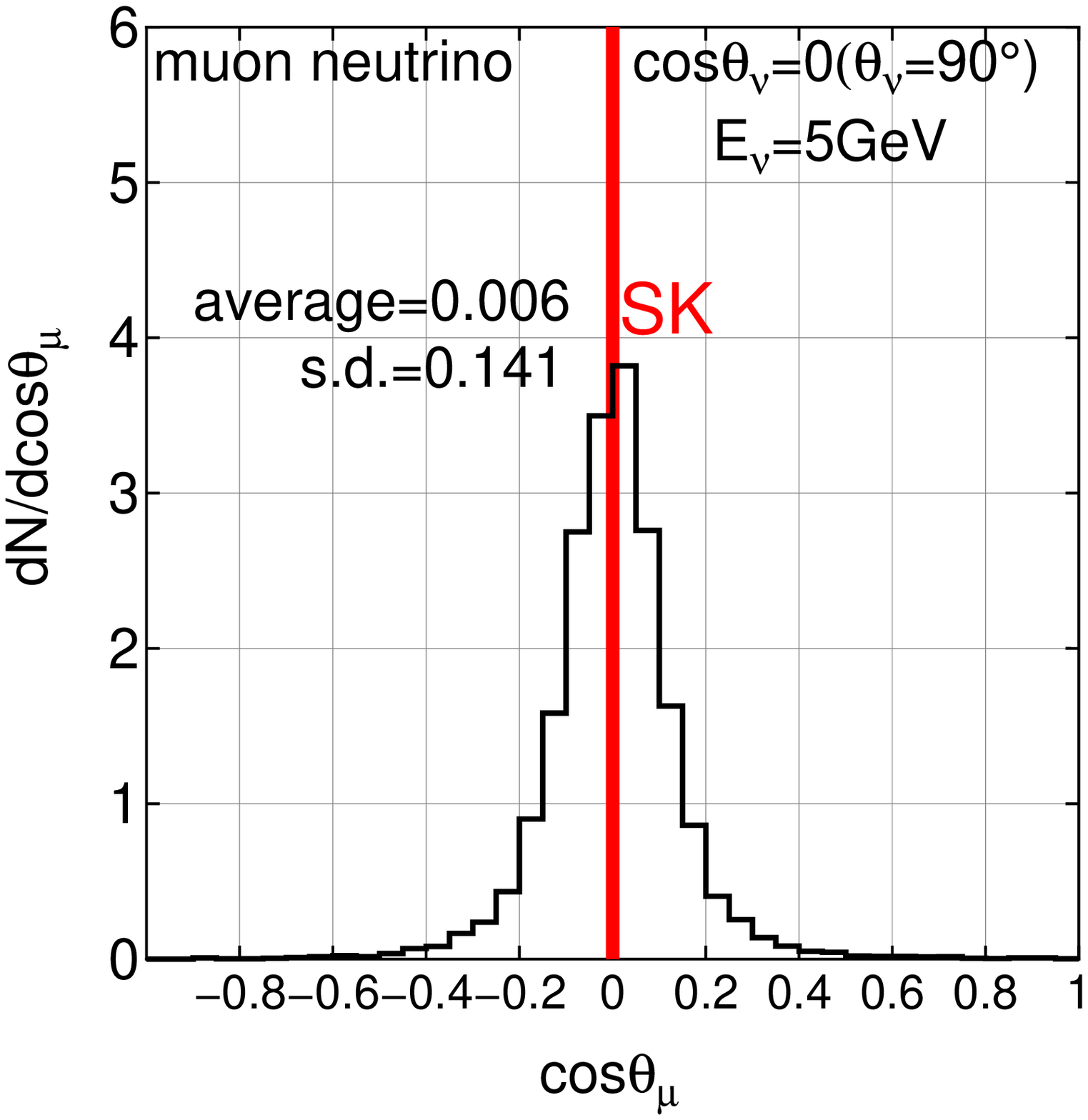}
  }
\caption{
\label{fig:8} 
Zenith angle distribution of the muon for the horizontally incident muon 
neutrino with 0.5~GeV, 1~GeV and 5~GeV, respectively. The sampling number 
is 10000 for each case.
SK stands for the corresponding ones under the SK assumption.
}
\end{center}
\vspace{0.5cm}
\hspace{2cm}(a)
\hspace{5cm}(b)
\hspace{5cm}(c)
\vspace{-0.5cm}
\begin{center}
\resizebox{\textwidth}{!}{%
  \includegraphics{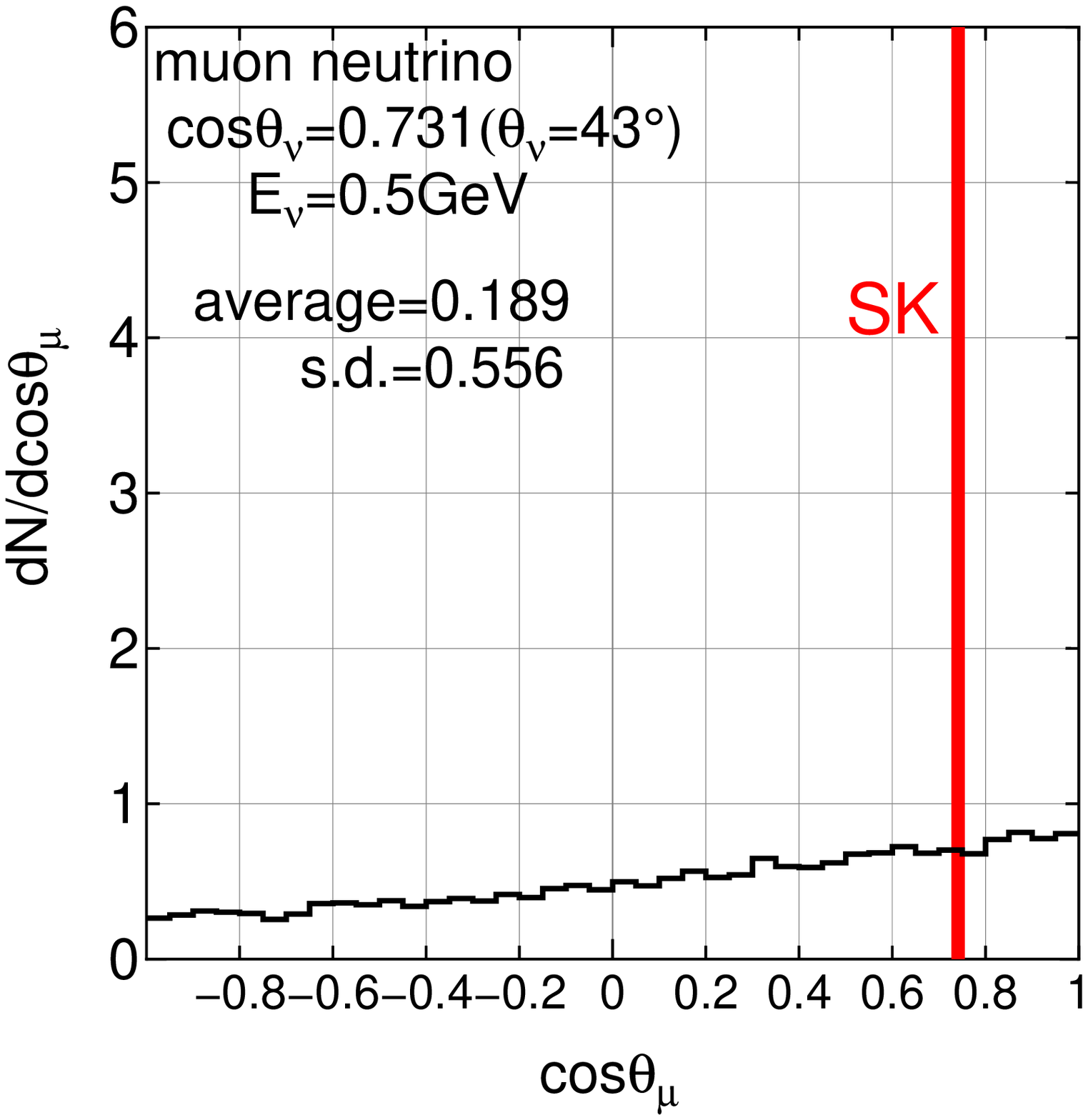}\hspace{1cm}
  \includegraphics{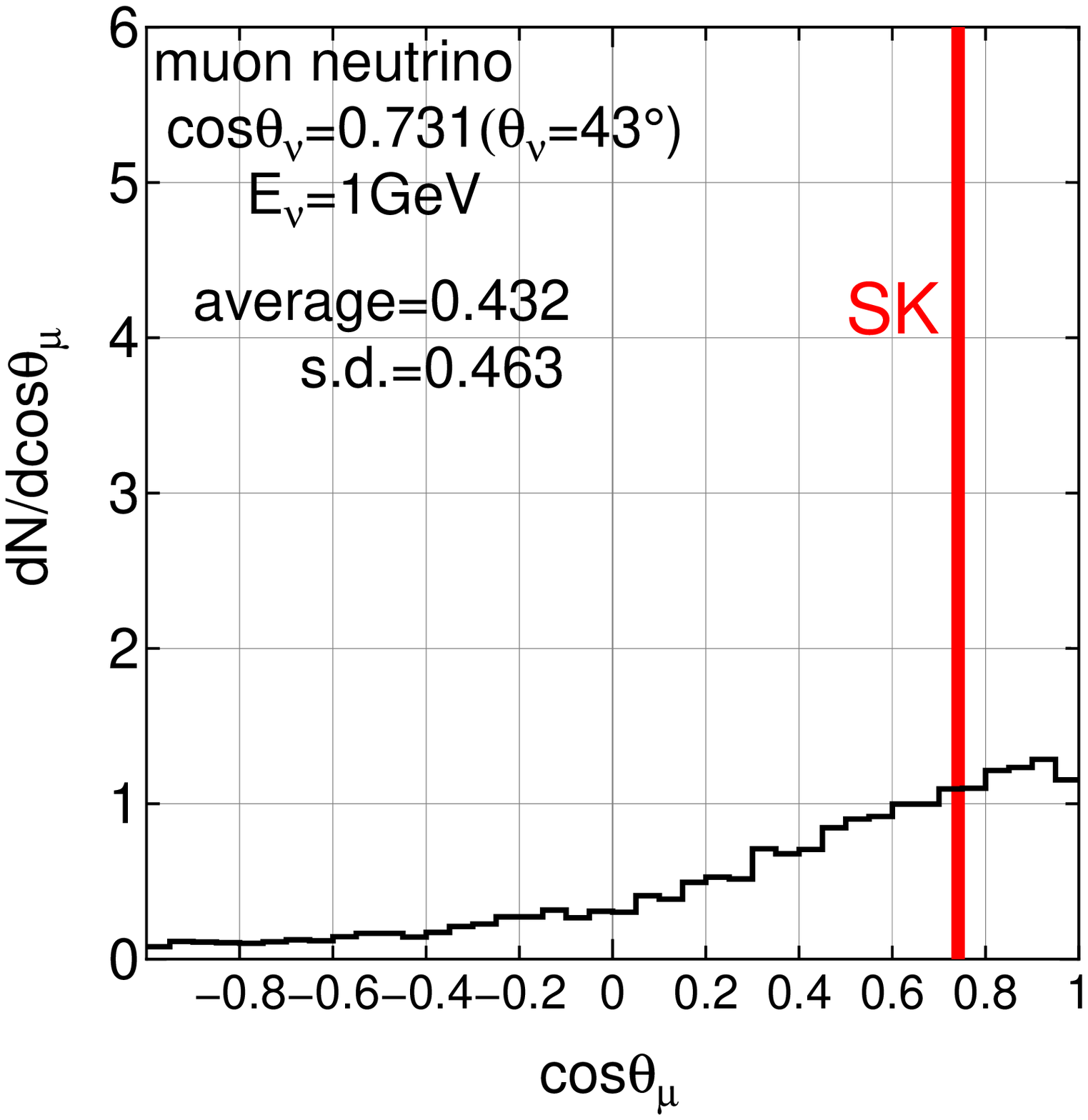}\hspace{1cm}
  \includegraphics{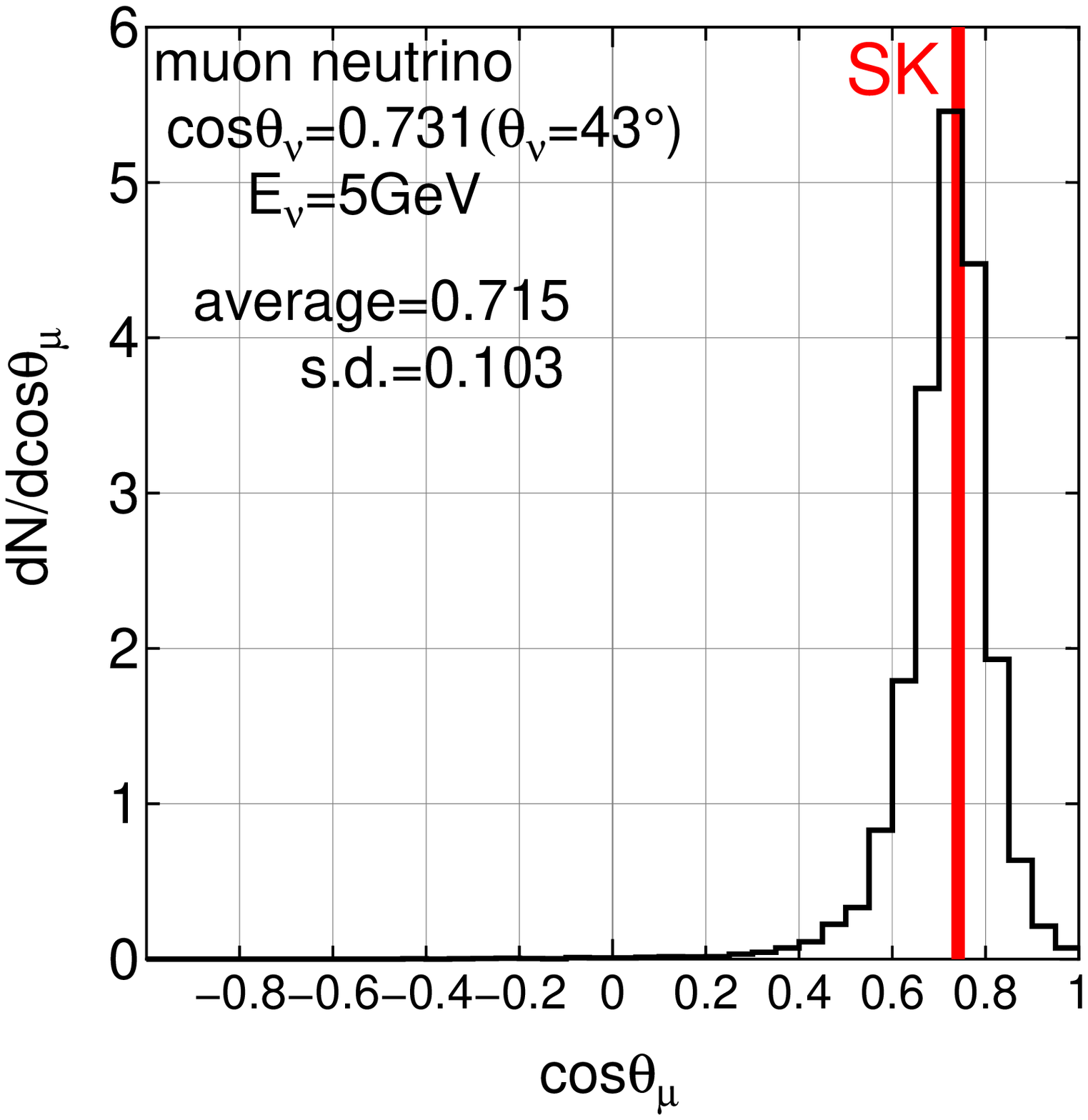}
  }
\caption{
\label{fig:9} 
Zenith angle distribution of the muon for the diagonally incident muon 
neutrino with 0.5~GeV, 1~GeV and 5~GeV, respectively. The sampling 
number is 10000 for each case.
SK stands for the corresponding ones under the SK assumption.
}
\end{center}
\end{figure*} 

\subsection{Zenith angle distribution of the emitted lepton 
for the different incidence of the neutrinos with different energies}
In Figures 7 to 9, we express Figs. 4 to 6 in a different way. 
We sum up muon events with different emitted energies for given 
zenith angles. As the result of it, we obtain frequency 
distribution of the neutrino events as  a function of 
$cos\theta_{\mu}$ for 
different incident directions and different incident energies of 
neutrinos.

In Figures~7(a) to 7(c), we give the zenith angle 
distributions of the emitted muons for the case of vertically incident 
neutrinos with different energies, say, $E_{\nu}=0.5$ GeV, $E_{\nu}=1$ GeV 
and $E_{\nu}=5$ GeV.

Comparing the case for 0.5 GeV with that for 5 GeV, we understand the big 
contrast between them as for the zenith angle distribution. The scattering 
angle of the emitted muon for 5 GeV neutrino is relatively small (See, 
Table 1) so that the emitted muons keep roughly the same direction as 
their 
original neutrinos. In this case, the effect of their azimuthal angle on 
the zenith angle is also small. However, in the case for 0.5 GeV which is  
the dominant energy for \textit{Fully Contained Events} 
and \textit{Partially Contained Events}
in the Superkamiokande, there is even 
a possibility for the emitted muon to be emitted in the backward direction 
due to the large angle scattering, the effect of which is enhanced by 
their azimuthal angle.

The most frequent occurrence in the backward direction of the emitted 
muon appears in the horizontally incident neutrino as shown in Figs. 8(a) 
to 8(c). In this case, the zenith angle distribution of the emitted muon 
should be symmetrical to the horizontal direction. Comparing the case for 
5 GeV with those both for $\sim$0.5 GeV and for $\sim$1 GeV, even 1 GeV 
incident neutrinos lose almost the original sense of the incidence if we 
measure it by the zenith angle of the emitted muon. Figs. 9(a) to 9(c) for 
the diagonally incident neutrino tell us that the situation for diagonal 
cases lies between the case for the vertically incident neutrino and that 
for horizontally incident one.
  SK in the figures denotes {\it the SK assumption on the direction} of 
incident neutrinos. 
From the Figures 7(a) to 9(c), it seems to be clear that the scattering 
angles of emitted muons influence their zenith angles, which is enhanced 
by their azimuthal angles, particularly more inclined directions of the 
incident neutrinos.

\begin{figure*}
\hspace{2cm}(a)
\hspace{5cm}(b)
\hspace{5cm}(c)
\vspace{-0.5cm}
\begin{center}
\resizebox{\textwidth}{!}{%
  \includegraphics{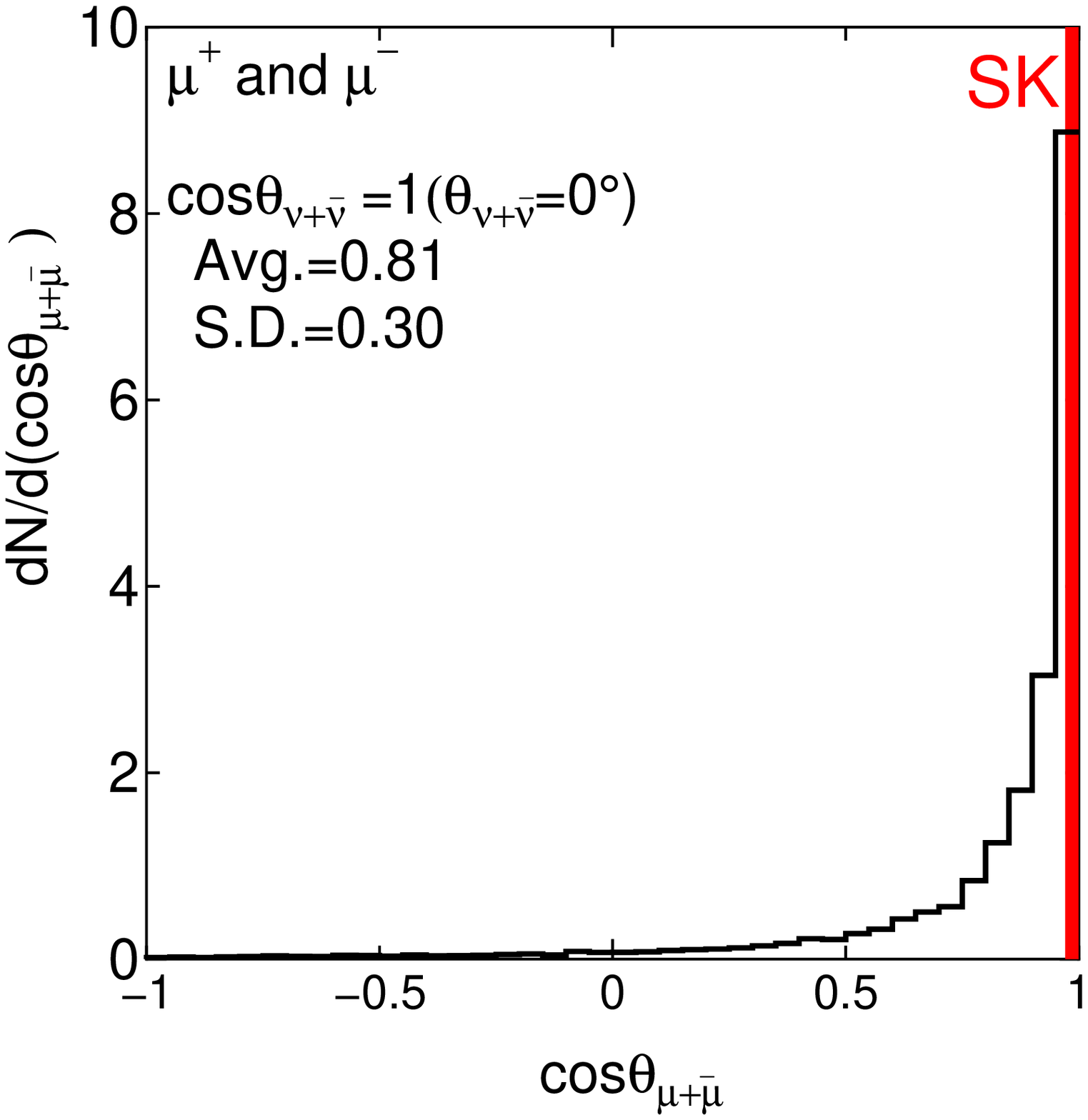}\hspace{1cm}
  \includegraphics{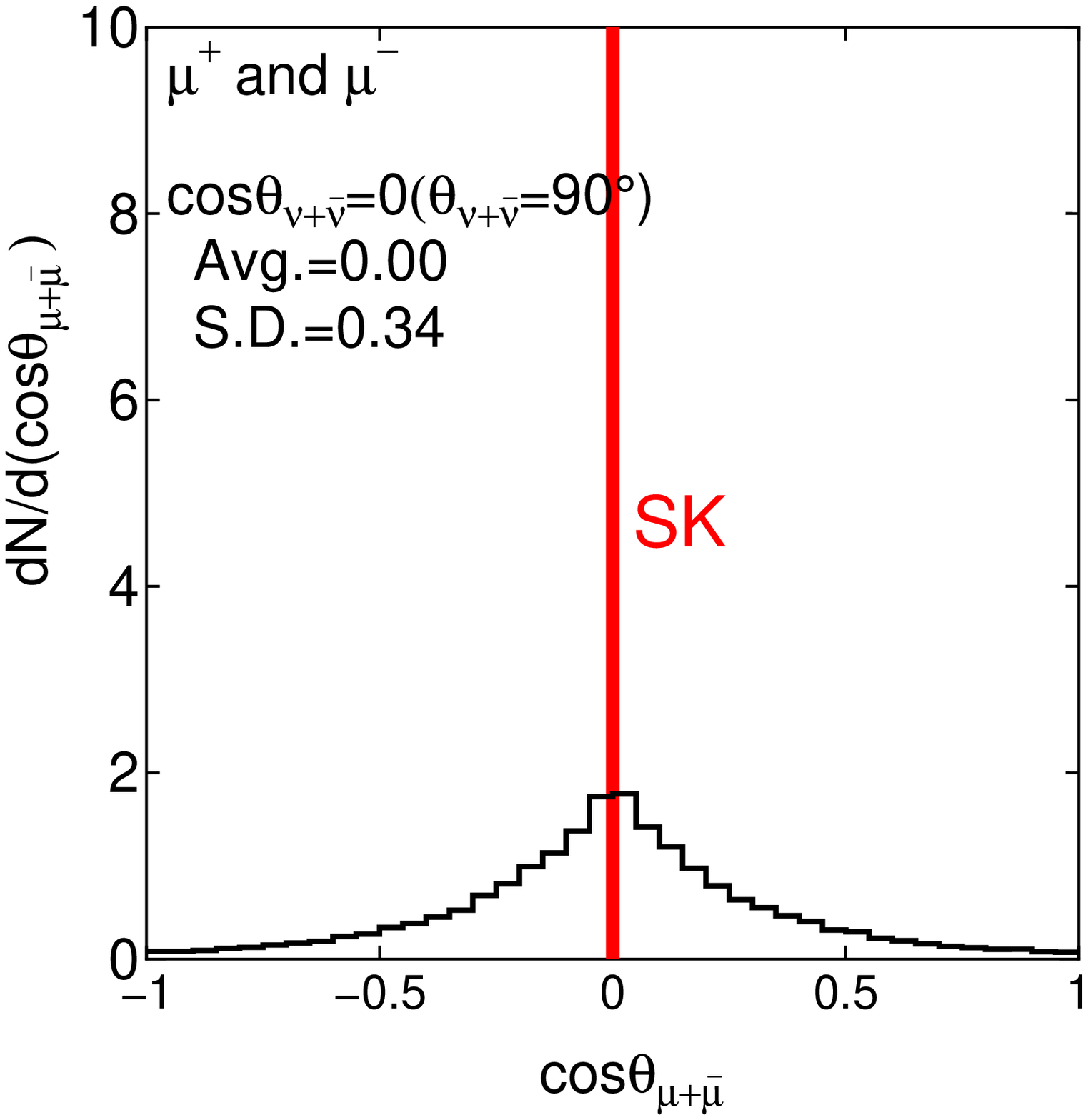}\hspace{1cm}
  \includegraphics{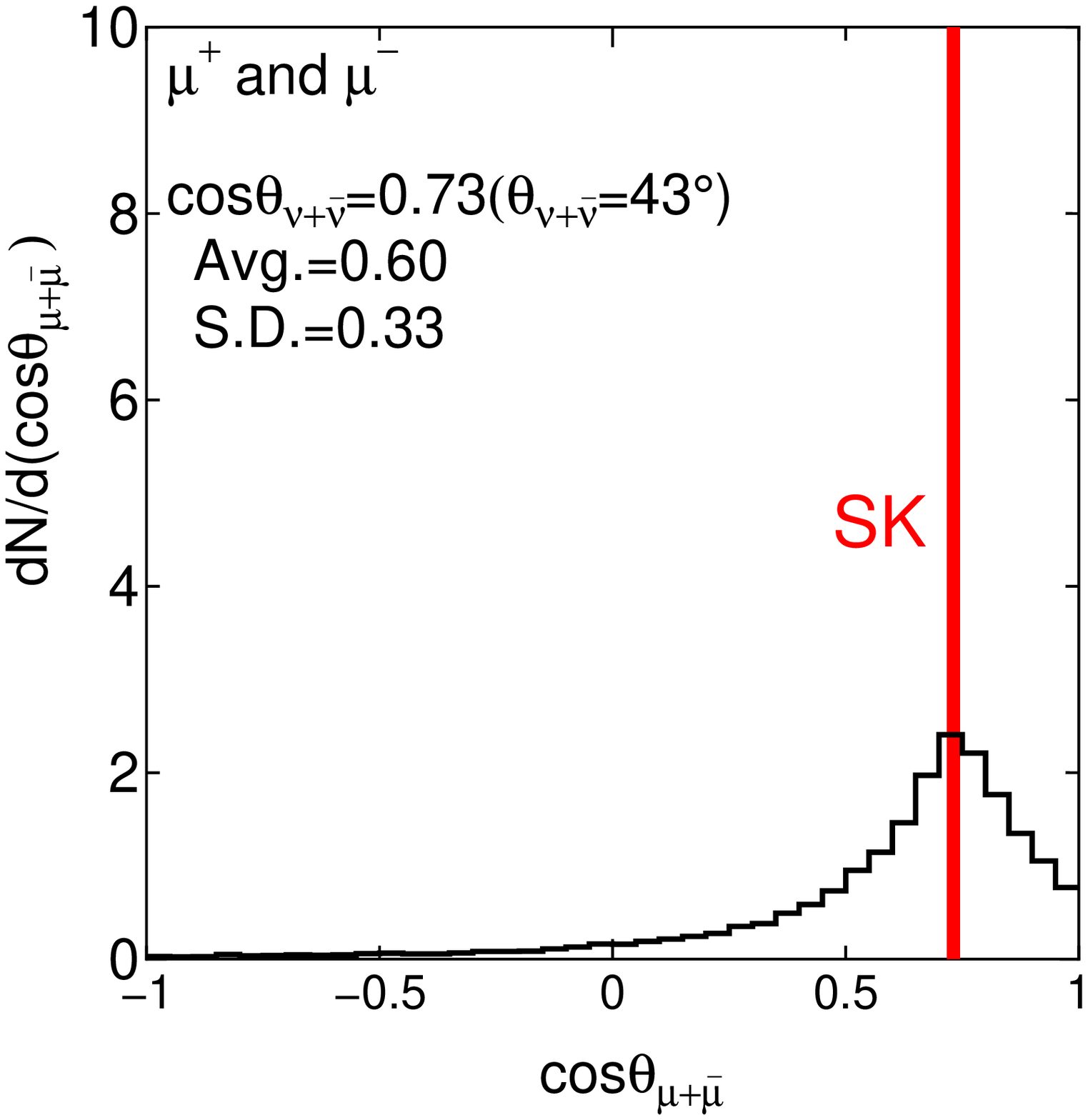}\hspace{1cm}
  }\\
 \end{center}
\caption{\label{fig:10} 
Zenith angle distribution of $\mu^-$ and $\mu^+$ for  $\nu$  and   
$\bar{\nu}$ for the incident neutrinos with the vertical, horizontal and 
diagonal directions, respectively (see Figure~3). 
The overall neutrino spectrum at 
Kamioka site is taken into account. The sampling number is 10000 for each 
case. SK stand for the corresponding ones under the SK assumption.
}
\end{figure*} 
%
\section{Zenith Angle Distribution of Fully Contained Events and 
 Partially Contained Events 
for a Given Zenith Angle of the Incident Neutrino, Taking Their Energy 
Spectrum into Account}

In the previous sections, we discuss the relation between the zenith angle 
distribution of the incident neutrino with a single energy and that of the 
emited muons produced by the neutrino for the different incident 
direction. In order to apply our motivation  around the uncertainty of 
\textit{the SK assumption on the direction} for \textit{Fully Contained 
Events} and \textit{Partially Contained Events}, we must consider the 
effect of the energy spectrum of the incident neutrino. The Monte Carlo 
simulation procedures for this purpose are given in the Appendix B.

In Fig. \ref{fig:10}, we give the zenith angle distributions of the sum of 
$\mu^+(\bar{\mu})$ and $\mu^-$ for a given zenith angle of 
$\bar{\nu}_{\bar{\mu}}$ and $\nu_{\mu}$, taking into account
different primary neutrino energy spectra for differnt directions
at Kamioka site.
SK in the figures denotes {\it the SK assumption on the direction}. 
From the figures, it seems to be clear that 
{\it the SK assumption on the direction} does not hold.
 Namely, we could 
conclude that the scattering angle of the emitted muons 
acompanied by 
their azimuthal angles influence their zenith angle distribution for 
the all directions. 


\section{Correlation between the Zenth Angle Distribution of the Incident 
Neutrinos and that of the Emitted Leptons}

Now, we extend the results for the definite zenith angle obtained in the 
previous sections to the case in which we consider the zenith angle 
distribution of the incident neutrinos totally.

Here, we examine the real correlation between ${\cos\theta_{\nu}}$ and 
${\cos{\theta}_{\mu}}$, by performing the exact Monte Carlo simulation. 
 

The detail for the simulation procedure is given in the Appendix C.

In order to obtain the zenith angle distribution of the emitted 
leptons for that of the incident neutrinos, we divide the cosine 
of the zenith angle of the incident neutrino into twenty regular 
intervals from $\cos{\theta_{\nu}}=0$ to $\cos{\theta_{\nu}}=1$.
For the given interval of $\cos{\theta_{\nu}}$, we carry out the 
exact Monte Carlo simulation, and obtain the cosine of the zenith 
angle of the emitted leptons.
 
Thus, for each interval of $\cos{\theta_{\nu}}$, we obtain the
corresponding zenith angle distribution of the emitted leptons. 
Then, we sum up these corresponding ones over all zenith angles of 
the incident neutrinos and we finally obtain the relation between 
the zenith angle distribution for the incident neutrinos and that 
for the emitted leptons. 

In a similar manner, we could obtain between $\cos{\theta_{\bar{\nu}}}$
and $\cos{\theta_{\bar{\mu}}}$ for anti-neutrinos. The situation 
for anti-\\
neutrinos is essentially the same as that for neutrinos.

\begin{figure*}
\begin{center}
\resizebox{0.45\textwidth}{!}{%
  \includegraphics{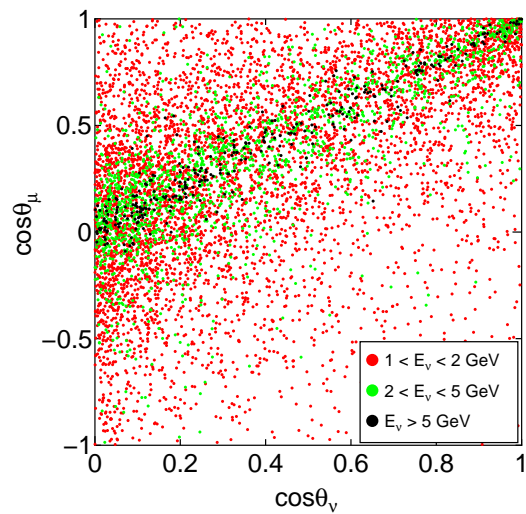}
  }
\caption{\label{fig:11} Correlation diagram between $\cos{\theta}_{\nu}$ and $\cos{\theta}_{\mu}$ 
for different neutrino energy regions.}
\end{center}
\end{figure*}

In Fig. \ref{fig:11} we classsify the correlation between 
${\cos\theta_{\nu}}$ and ${\cos{\theta}_{\mu}}$ according to the different 
energy range of the incident muon neutrinos. 
The ${\cos{\theta}_{\mu}}$ distribution along 
${\cos\theta_{\nu}}=1$, 
${\cos\theta_{\nu}}=0$ and
${\cos\theta_{\nu}}=0.73$ in Figure~11 correspond to Figure~10(a) (vertical), Figure~10(b)(horizontal) and Figure~10(c)
(diagonal), respectively.  
Looking the ${\cos\theta_{\mu}}$ distribution for the fixed
${\cos\theta_{\nu}}$    in the Figure~11,
it is well understood that the   ${\cos\theta_{\mu}}$ distribution
around ${\cos\theta_{\mu}}\approx {\cos\theta_{\nu}}$
spreads wider as  ${\cos\theta_{\nu}}$    decreases.
This is due to the effect of scattering angle
enhanced by the azimuthal angle (see Figures~3 and 15, also).

\begin{figure*}
\begin{center}
\resizebox{0.9\textwidth}{!}{%
\includegraphics{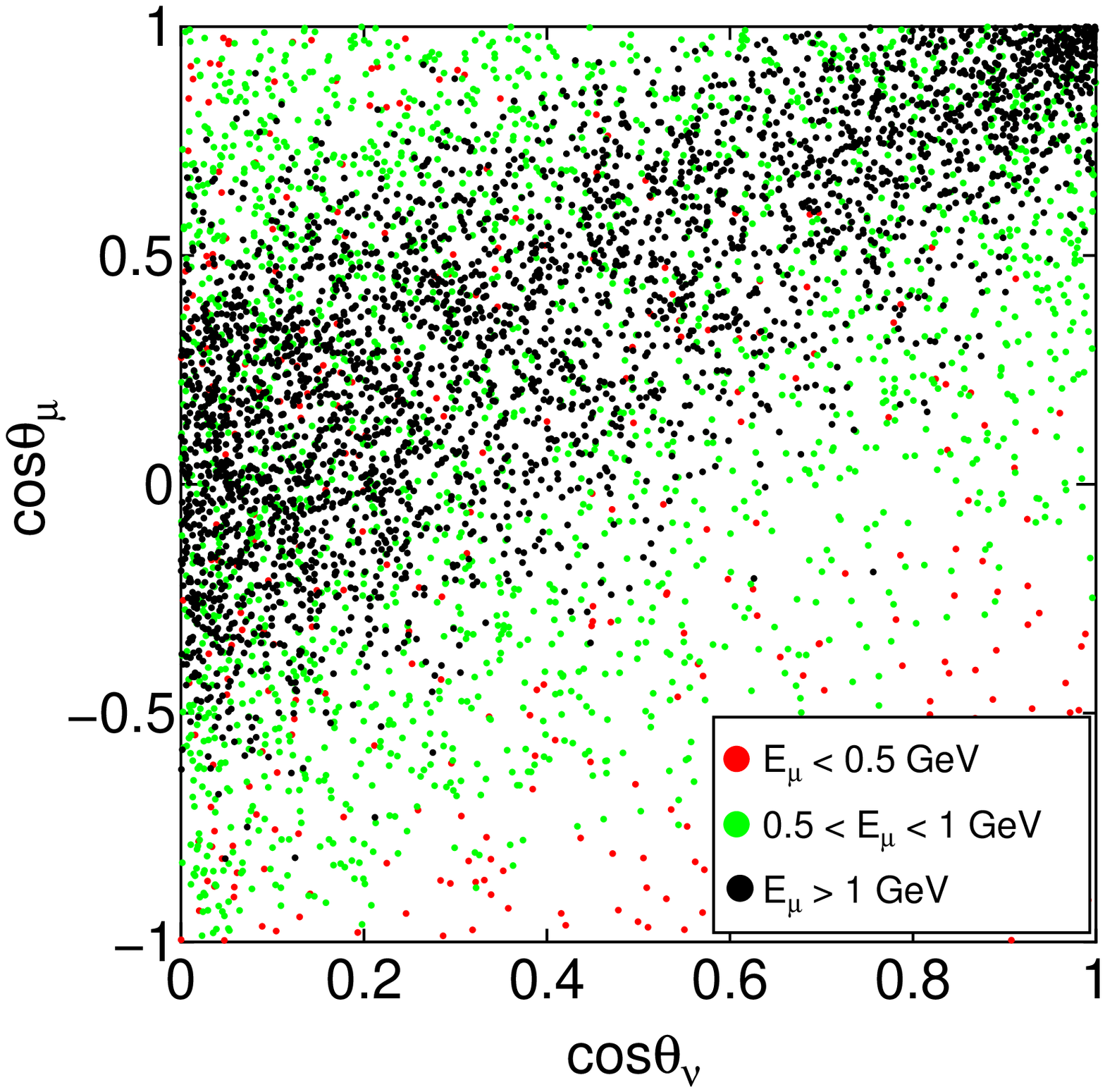}
\hspace{1cm}
\includegraphics{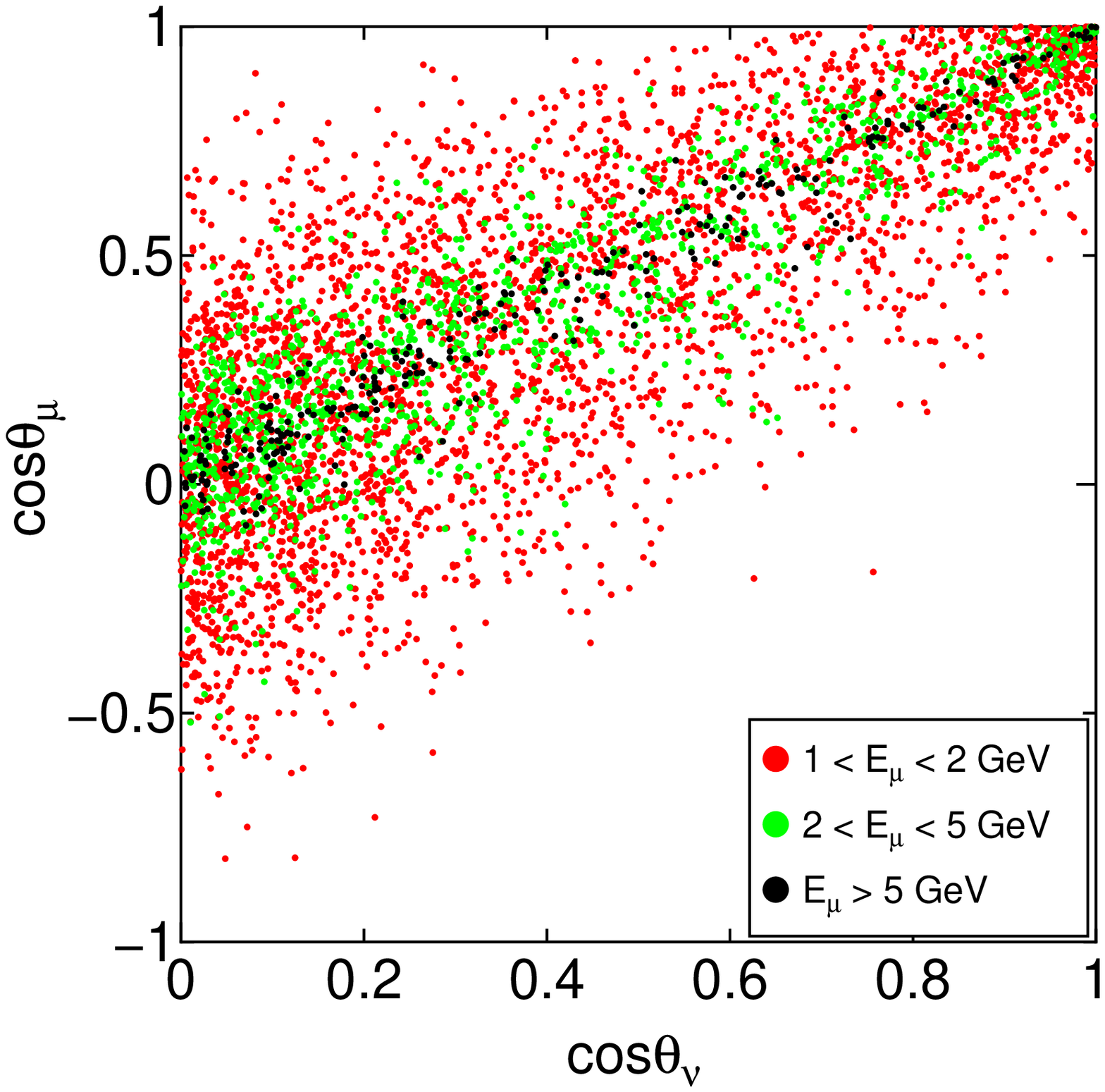}
}
\caption{\label{fig:12} Correlation diagrams between ${\cos\theta_{\nu}}$ 
and $\cos{\theta}_{\mu}$ 
for different muon energy ranges.}
\end{center}
\end{figure*}

In Fig. \ref{fig:12}, we classify the correlation between 
${\cos\theta_{\nu}}$ and ${\cos\theta_{\mu}}$ according to the different 
energy range of $E_{\mu}$. 
It is clear from Figure~12 that 
(a){\it the SK assumption on the direction} 
(${\cos\theta_{\mu}}\approx {\cos\theta_{\nu}}$) 
does hold surely $E_{\mu} >$ 5 GeV,
(b)this assumption does hold rougly $E_{\mu} >$ 2 GeV.
However, as $E_{\mu}$ decreases, this relation 
becomes incorrect.
 In the energy range of 0.5~$< E_{\mu} <$ 1~GeV where 
neutrino events
in the Super-Kamiokande detector mostly exist, 
it does not hold completely. 


Thus, it could be surely concluded from Fig. \ref{fig:11} and Fig. 
\ref{fig:12}  that {\it the SK assumption on the direction}
 does not hold as 
a good estimator for the determination of the directions of the 
incident neutrinos even if statistically.
 
Finally, we examine the relation between the zenith angle distribution
 for upward $\nu_{\mu}$ and $\bar{\nu}_{\mu}$ 
and the corresponding zenith angle distribution for $\mu$ and 
$\bar{\mu}$ in the case of no oscillation. By perfoming the procedure 
described in Appendix C, we obtain a pair of 
($\cos{\theta_\mu}$, $E_\mu$) from a sampling of 
($\cos{\theta_\nu}$, $E_\nu$).
In Figure~13, we show the relation between the upward incident neutrino zenith angle distribution and the emitted muon ones thus obtained for the null oscillation. 
 If we guarantee {\it the SK assumption on the direction}, 
the emitted muon zenith angle distribution is expressed approximately as 
the incident neutrino zenith angle distribution. However, the really 
simulated muon zenith angle distribution is quite different from the 
incident neutrino zenith angle distribution. This is the reason why 
{\it the SK assumption on the direction}
 does not hold even if statistically.
 Furthermore, it should be noticed from the figure that the existence 
of the downward muons from the upward neutrinos could not be neglected 
which is enhanced by the azimuthal angles in addition to the 
backscattering. 
If the neutrino oscillation does not exist,
 such downward muons do not bring about problems, because in this 
case the zenith angle distribution of the downward neutrino is 
symmetrical to that of upward neutrino. Then, the 
"inverted leptons" cancel out totally. However, in the case of the presence 
of the neutrino oscillation, such cancellation does not occur. 
Namely, the determination of the direction of the incident neutrinos 
estimated by the emitted muon should be more carefully treated.

\begin{figure}
\begin{center}
\resizebox{0.45\textwidth}{!}{%
  \includegraphics{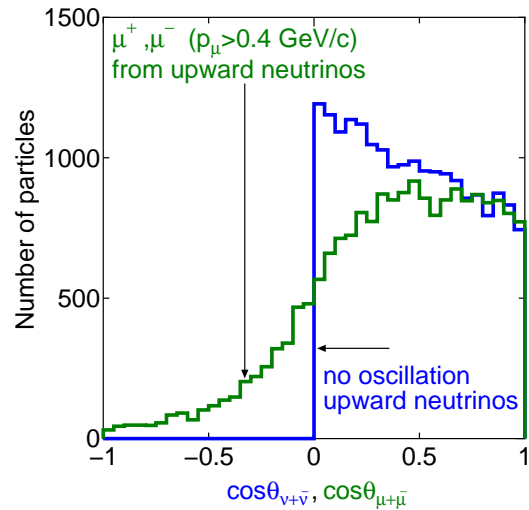}
  }
\caption{\label{fig:13} The relation between the zenith angle distribution of the incident neutrino and
corresponding ones of the emitted lepton}
\end{center}
\end{figure}

\section{Conclusions}
We have shown the invalidity of 
{\it the SK assumption on the direction} for the analysis of 
the Single Ring $\mu$ Events among 
{\it Fully Contained Events} which are solid and free from
the interpretation on their interaction types.
Super-Kamiokande Collaboration have examined 
all possible neutrino events for the neutrino oscillation
 problem, say, Sub-GeV e-like Events, Multi-GeV e-like Events,
 Sub-GeV $\mu$-like Events, Multi-GeV $\mu$-like Events, 
multi-ring Sub GeV $\mu$-like Events, multi-ring Multi-GeV Events, 
PC Events, {\it Upward Stopping Muon Events} and 
{\it Upward Through Going Muon Events}
 (see, \cite{Ashie}), 
and  all possible neutrino events provide the same neutrino oscillation 
parameters with the same precision accidentally. 
However, different types of neutrino events have different structures 
which are accompanied by different experimental uncertainties. Therefore, 
we could not readily believe such unified conclusions, taking into 
consideration the different experimental qualities in the different types 
of events. 
The most clear cut events among the SK events are one ring electron like
 events and muon-like events among {\it Fully Contained Events}. 
In these events, apart from the numerical uncertainties, one could 
essentially
discriminate electron from muon (see footnote 1), and
we decide the directions of the leptons as well as their energies,
 because we know the interaction points as well as their end points 
in the detector. Therefore, it is more desirable that one analyze 
the single-ring electron-like events and single-ring muon-like events 
among {\it Fully Contained Events} exclusively not being tempted by the 
increase of statistics.
If one really finds the solid evidence for the neutrino oscillation 
in the most clear cut events, such as
{\it Single Ring $\mu$ Events}, then one could find some corroboration 
for the neutrino oscillation even in the worst quality of 
experimental data, such as 
{\it Upward Stopping Muon Events} and 
{\it Upward Through Going Muon Events}.  

In the subsequent paper (Part~2), we apply our present result to the
 analysis of the $L/E$ distribution for 
{\it Single Ring $\mu$ Events} 
due to QEL among {\it Fully Contained Events}  
which may lead the direct observation of the neutrino oscillation,
if really exists, 
due to the simplicity of the events concerned. 
Also, we are now under preparation the 
third paper (Part~3) in which we examine the zenith angle 
distribution for the same type of the events mentioned above.

\newpage
In the following Appendices we give the concrete Monte Carlo Simulations, namely, the details of our \textit{Time Sequential Simulation}.  
\appendix

\section{\appendixname:{} Monte Carlo Procedure for the Decision of 
Emitted Energies of the Leptons and Their Direction Cosines }
\setcounter {equation} {0}
\def\theequation{\Alph{section}\textperiodcentered\arabic{equation}}

Here, we give the Monte Carlo Simulation procedure for obtaining the 
energy and its  direction cosines,
$(l_{r},m_{r},n_{r})$, of the emitted lepton in QEL for a given energy and 
its direction cosines, $(l,m,n)$, of the incident neutrino. 

The relation among $Q^2$, $E_{\nu+\bar{\nu}}$, the energy of the incident 
neutrino, $E_{\ell}$, the energy of the emitted lepton (muon or electron 
or their anti-particles) and $\theta_{\rm s}$, the scattering angle of the 
emitted lepton, is given as
      \begin{equation}
         Q^2 = 2E_{\nu(\bar{\nu})}E_{\ell(\bar{\ell})}(1-{\rm cos}\theta_{\rm s}).
\label{eqn:a1}  
      \end{equation}
\noindent Also, the energy of the emitted lepton is given by
      \begin{equation}
         E_{\ell(\bar{\ell})} = E_{\nu(\bar{\nu})} - \frac{Q^2}{2M}.
\label{eqn:a2}  
      \end{equation}

\begin{figure}
\begin{center}
\resizebox{0.45\textwidth}{!}{%
  \includegraphics{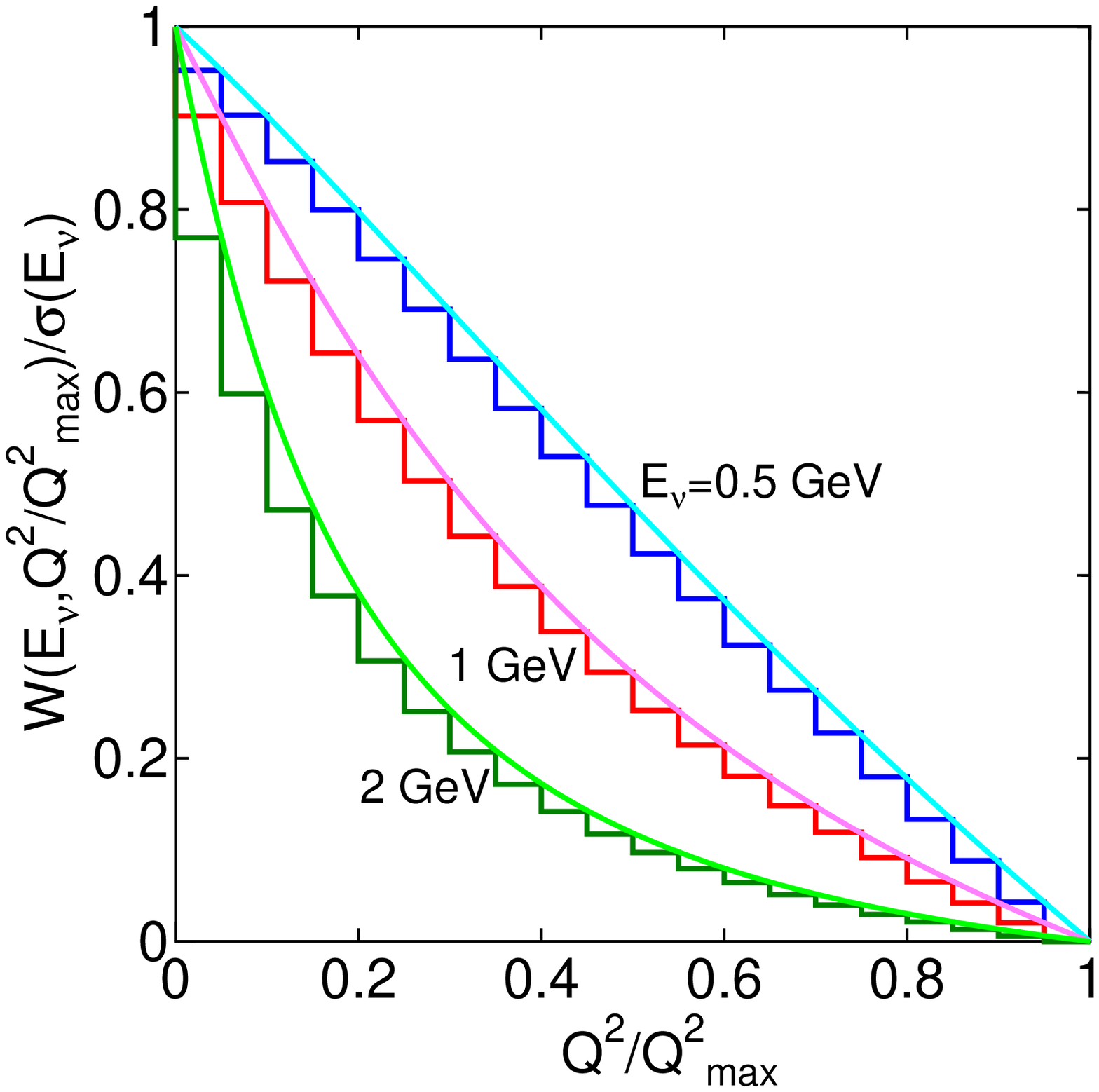}
  }
\end{center}
\caption{\label{fig:a1} The reappearance of the probability function for 
QEL cross section.
Histograms are  sampling results, while the curves  concerned are 
theoretical
ones for given incident energies.
}
\end{figure}

\noindent {\bf Procedure 1}\\
\noindent
We decide  $Q^2$ from the probability function for the differential cross 
section with a given $E_{\nu(\bar{\nu})}$ (Eq. (\ref{eqn:2}) in the text) 
by using the uniform random number, ${\xi}$,  between (0,1) in the 
following\\
  \begin{equation}
    \xi = \int_{Q_{\rm min}^2}^{Q^2}P_{\ell(\bar{\ell})}(E_{\nu(\bar{\nu})},Q^2)
                             {\rm d}Q^2,
\label{eqn:a3}  
  \end{equation}
\noindent where
  \begin{eqnarray}
\lefteqn{     P_{\ell(\bar{\ell})}(E_{\nu(\bar{\nu})},Q^2) =} \nonumber \\
&&  \frac{ {\rm d}\sigma_{\ell(\bar{\ell})}(E_{\nu(\bar{\nu})},Q^2) }{{\rm d}Q^2} 
                     \Bigg /\!\!\!\!
      \int_{Q_{\rm min}^2}^{Q_{\rm max}^2} 
      \frac{ {\rm d}\sigma_{\ell(\bar{\ell})}(E_{\nu(\bar{\nu})},Q^2) }{{\rm d}Q^2} 
             {\rm d}Q^2 . \nonumber \\
&&
\label{eqn:a4}  
   \end{eqnarray}
\\
\noindent From Eq. (A$\cdot$1), we obtain $Q^2$ in histograms together 
with the corresponding theoretical curve in Fig. \ref{fig:a1}. The 
agreement between the sampling data and the theoretical curve is 
excellent, which shows the validity of the utlized  procedure in Eq. (A
$\cdot$3) is right. \\

\noindent {\bf Procedure 2}\\
\noindent
We obtain $E_{\ell(\bar{\ell})}$ from Eq. (A$\cdot$2) for  the given 
$E_{\nu(\bar{\nu})}$ and $Q^2$ thus decided in the Procedure 1.\\

\noindent {\bf Procedure 3}\\
\noindent
We obtain $\cos{\theta_{\rm s}}$, cosine of the the scattering angle of 
the emitted lepton, for $E_{\ell(\bar{\ell})}$ thus decided in the 
Procedure 2 from Eq. (A$\cdot$1) .\\

\noindent {\bf Procedure 4}\\
\noindent
We decide $\phi$, the azimuthal angle of the scattering lepton, which is obtained from\\
  \begin{equation}
       \phi = 2\pi\xi.                 
\label{eqn:a5}  
  \end{equation}

\noindent Here, $\xi$ is a uniform random number of the range (0, 1). \\
As explained schematically in the text(see Fig. {\bf \ref{fig:3}} in the 
text),  we must take account of the effect due to the azimuthal angle 
$\phi$ in the QEL to obtain the zenith angle distribution of both 
{\it Fully Contained Events} and {\it Partially Contained Events} 
correctly.\\  

\noindent {\bf Procedure 5}\\
\noindent
The relation between direction cosines of the incident neutrinos, 
$(\ell_{\nu(\bar{\nu})}, m_{\nu(\bar{\nu})}, n_{\nu(\bar{\nu})} )$, and 
those of the corresponding emitted lepton, $(\ell_{\rm r}, m_{\rm r}, 
n_{\rm r})$, for a certain $\theta_{\rm s}$ and $\phi$ is given as \\

\begin{equation}
\left(
         \begin{array}{c}
             \ell_{\rm r} \\
             m_{\rm r} \\
             n_{\rm r}
         \end{array}
       \right)
           =
       \left(
         \begin{array}{ccc}
            \displaystyle \frac{\ell n}{\sqrt{\ell^2+m^2}} & 
            -\displaystyle 
            \frac{m}{\sqrt{\ell^2+m^2}}        & \ell_{\nu(\bar{\nu})} \\
            \displaystyle \frac{mn}{\sqrt{\ell^2+m^2}} & \displaystyle 
            \frac{\ell}{\sqrt{\ell^2+m^2}}     & m_{\nu(\bar{\nu})}    \\
                        -\sqrt{\ell^2+m^2} & 0 & n_{\nu(\bar{\nu})}
         \end{array}
       \right)
       \left(
          \begin{array}{c}
            {\rm sin}\theta_{\rm s}{\rm cos}\phi \\
            {\rm sin}\theta_{\rm s}{\rm sin}\phi \\
            {\rm cos}\theta_{\rm s},
          \end{array}
       \right),
\label{eqn:a6}
\end{equation}
\noindent where $n_{\nu(\bar{\nu})}={\rm cos}\theta_{\nu(\bar{\nu})}$, and 
$n_{\rm r}={\rm cos}\theta_{\ell}$. 
Here, $\theta_{\ell}$ is the zenith angle of the emitted lepton. \\

\begin{figure}
\begin{center}
\resizebox{0.5\textwidth}{!}{%
  \includegraphics{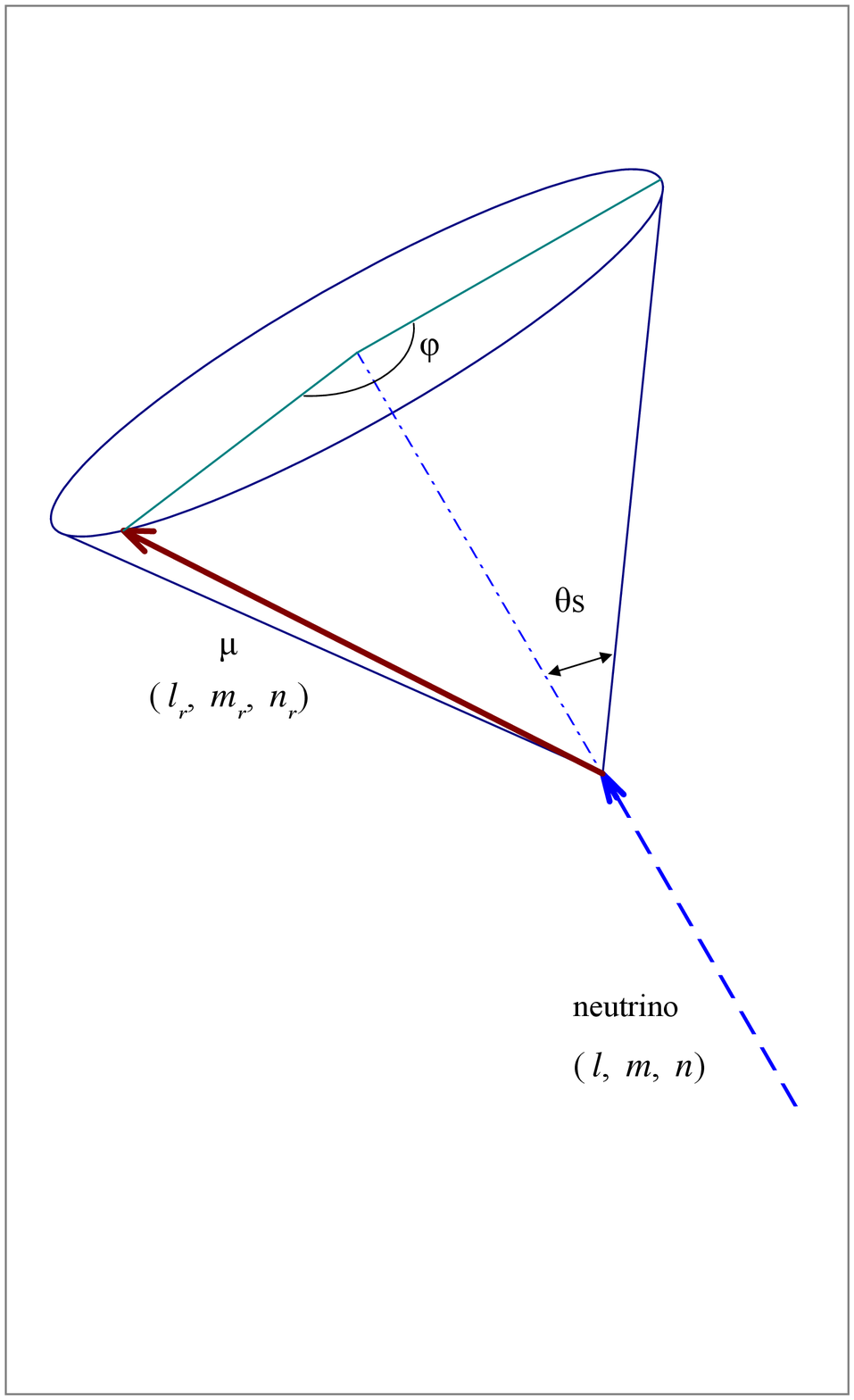}
  }
\end{center}
\caption{\label{fig:14} The relation between the direction cosine of the 
incident neutrino and that of the emitted charged lepton.}
\end{figure} 

The Monte Carlo procedure for the determination of $\theta_{\ell}$ of the 
emitted lepton for the parent (anti-)neutrino with given 
$\theta_{\nu(\bar{\nu})}$ and $E_{\nu(\bar{\nu})}$ involves the following 
steps:\\

We obtain $(\ell_r, m_r, n_r)$ by using Eq. (\ref{eqn:a6}). The $n_r$ is 
the cosine of the zenith angle of the emitted lepton which should be 
contrasted to $n_{\nu}$, that of the incident neutrino.
\\
Repeating the procedures 1 to 5 just mentioned above, we obtain the zenith 
angle distribution of the emitted leptons for a given zenth angle of the 
incident neutrino with a definite energy. \\

In the SK analysis,  instead of Eq. (\ref{eqn:a6}), they assume \\ $n_r = 
n_{\nu(\bar{\nu})} $ 
uniquely for ${E_{\mu(\bar{\mu})}} \geq$ 400 MeV.\\

\section{\appendixname:{ Monte Carlo Procedure to Obtain the Zenith Angle 
of the Emitted Lepton for a Given Zentith Angle of the Incident Neutrino}}
 \setcounter {equation}{0}
 \def\theequation{\Alph{section}\textperiodcentered\arabic{equation}}

  The present simulation procedure for a given zenith angle of the 
incident neutrino starts from the atmospheric neutrino spectrum at the 
opposite site of the Earth to the SK detector. We define, $N_{\rm 
int}(E_{\nu(\bar{\nu})},t,{\rm cos}\theta_{\nu(\bar{\nu})})$, the 
interaction neutrino spectrum at the depth $t$ from the SK detector in the 
following way
   \begin{eqnarray}
    \lefteqn{  N_{\rm int}(E_{\nu(\bar{\nu})},t,{\cos}\theta_{\nu(\bar{\nu
})}) =N_{\rm sp}(E_{\nu(\bar{\nu})},\cos\theta_{\nu(\bar{\nu})}) \times } 
\nonumber \\
&&  \Bigg(1-\frac{{\rm d}t}{\lambda_1(E_{\nu(\bar{\nu})},t_1,\rho_1)} 
\Bigg)  \times \cdots \times \Bigg(1-\frac{{\rm d}t}{\lambda_n(E_{\nu(\bar{
\nu})},t_n,\rho_n)} \Bigg).\nonumber \\
&& 
\label{eqn:b1}
   \end{eqnarray}

Here, $N_{\rm sp}(E_{\nu(\bar{\nu})},\cos\theta_{\nu(\bar{\nu})})$ is the 
atmospheric (anti-)  neutrino spectrum for the zenith angle at the 
opposite surface of the Earth.

%

Here $\lambda_i(E_{\nu(\bar{\nu})},t_i,\rho_i)$ denotes the mean free path 
of the neutrino (anti neutrino) with the energy $E_{\nu(\bar{\nu})}$ 
due to QEL at the distance, $t_i$, from the opposite surface of the Earth 
inside whose density is $\rho_i$. 


The procedures of the Monte Carlo Simulation for the incident 
neutrino(anti neutrino) with a given energy, $E_{\nu(\bar{\nu})}$, whose 
incident direction is expressde by $(l,m,n)$ are as follows.\\

\noindent {\bf Procedure A}\\
\noindent
For the given zenith angle of the incident neutrino, 
${\theta_{\nu(\bar{\nu})}}$, we formulate, $N_{\rm pro}( 
E_{\nu(\bar{\nu})},t,\cos\theta_{\nu(\bar{\nu})}){\rm d}E_{\nu(\bar{\nu})}$
, the production function for the neutrino flux to produce leptons at the 
Kamioka site in the following
   \begin{eqnarray}
\lefteqn{N_{\rm pro}( E_{\nu(\bar{\nu})},t,\cos\theta_{\nu(\bar{\nu})}){\rm
 d}E_{\nu(\bar{\nu})} } \nonumber \\
&&=
 \sigma_{\ell(\bar{\ell})}(E_{\nu(\bar{\nu})}) N_{\rm int}(E_{\nu(\bar{\nu
})},t,{\rm cos}\theta_{\nu(\bar{\nu})}){\rm d}E_{\nu(\bar{\nu})},
\label{eqn:b2}
   \end{eqnarray}
%
\noindent where
  \begin{equation}
     \displaystyle
      \sigma_{\ell(\bar{\ell})}(E_{\nu(\bar{\nu})}) = \int^{Q_{\rm 
max}^2}_{Q_{\rm min}^2}  \frac{ {\rm d}\sigma_{\ell(\bar{\ell})}(E_{
\nu(\bar{\nu})},Q^2)}{{\rm d}Q^2}{\rm d}Q^2.
\label{eqn:b3}
  \end{equation}

\noindent Each differential cross section above is given in Eq. (\ref
{eqn:2}) in the text.\\
Utilizing, $\xi$, the uniform random number between (0,1), 
we determine $E_{\nu(\bar{\nu})}$, the energy of the incident neutrino 
in the following sampling procedure\\
    \begin{equation}
       \xi = \int_{E_{\nu(\bar{\nu}),{\rm min}}}^{E_{\nu(\bar{\nu})}}
             P_d(E_{\nu(\bar{\nu})},t,\cos\theta_{\nu(\bar{\nu})})
{\rm d}E_{\nu(\bar{\nu})},
    \end{equation}
where
\begin{eqnarray}
\lefteqn{
        P_d(E_{\nu(\bar{\nu})},t,\cos{\theta}_{\nu(\bar{\nu})}){\rm d}E_{
\nu(\bar{\nu})} 
        } \nonumber\\
&=& 
\frac{
N_{pro}( E_{\nu(\bar{\nu})},t,\cos{\theta}_{\nu(\bar{\nu})}){\rm d}E_{
\nu(\bar{\nu})} 
}
{ \displaystyle \int_{E_{\nu(\bar{\nu}),{\rm min}}}^{E_{\nu(\bar{\nu}),{\rm
 max}}} 
                       N_{pro}( E_{\nu(\bar{\nu})},t,\cos{\theta}_{\nu(\bar
{\nu})}){\rm d}E_{\nu(\bar{\nu})} 
} .
\end{eqnarray}

In our Monte Carlo procedure, \\ 
 the reproduction of, 
$P_d(E_{\nu(\bar{\nu})},t,\cos\theta_{\nu(\bar{\nu})}){\rm 
d}E_{\nu(\bar{\nu})}$, 
the normalized differential neutrino interaction probability function, is 
confirmed in the same way as in Eq. (A$\cdot$4). 
\\
\\
%
%
%
%
 
\noindent {\bf Procedure B}\\
\noindent
For the (anti-)neutrino concerned with the energy of $E_{\nu(\bar{\nu})}$, 
we sample $Q^2$ utlizing $\xi_{3}$, the uniform random number between 
(0,1). The Procedure B is exactly the same as in the Procedure 1 in the 
Appendix A. \\

\noindent {\bf Procedure C}\\
\noindent
We decide, ${\theta_{\rm s}}$, the scattering angle of the emitted lepton 
for given $E_{\nu(\bar{\nu})}$ and $Q^2$. The procedure C is exactly the 
same as in the combination of Procedures 2 and 3 in the 
 Appendix A. \\

\noindent {\bf Procedure D}\\
\noindent
We randomly sample the azimuthal angle of the charged lepton concerned. 
The Procedure D is exactly the same as in the Procedure 4 in the Appendix 
A. \\
 
\noindent {\bf Procedure E}\\
\noindent
We decide the direction cosine of the charged lepton concerned. The 
Procedure E is exactly the same as in the Procedure 5 in the  Appendix A.\\

 We repeat Procedures A to E until we reach the desired trial number. \\
\\

\section{\appendixname:{ }Correlation between the Zenith Angles of the 
Incident Neutrinos and Those of the Emitted Leptons}
\setcounter{equation}{0}
\def\theequation{\Alph{section}\textperiodcentered\arabic{equation}}

\noindent {\bf Procedure A}\\
By using, $N_{\rm pro}( 
E_{\nu(\bar{\nu})},t,\cos\theta_{\nu(\bar{\nu})}){\rm d}E_{\nu(\bar{\nu})}$,
which is defined 
in Eq. (\ref{eqn:b2}), 
   
\noindent we define the spectrum for $\cos\theta_{\nu(\bar{\nu})}$  in the 
following.

\begin{eqnarray}
\lefteqn{      I(\cos\theta_{\nu(\bar{\nu})}){\rm d}(\cos\theta_{\nu(\bar{
\nu})}) = } \nonumber \\
&&      {\rm d}(\cos\theta_{\nu(\bar{\nu})})
       \int_{E_{\nu(\bar{\nu}),{\rm min}}}^{E_{\nu(\bar{\nu}),{\rm max}}}
 \!\!\!\!\!\!\!\!\!\!\!\!\!\!\!\!N_{\rm pro}( E_{\nu(\bar{\nu})},t,
\cos\theta_{\nu(\bar{\nu})}){\rm d}E_{\nu(\bar{\nu})}.
\label{eqn:c1}
\end{eqnarray}

\noindent By using Eq. (\ref{eqn:c2}) and $\xi$, a sampled uniform random 
number 
between (0,1), then we could determine $\cos\theta_{\nu(\bar{\nu})}$
from the following equation
    \begin{equation}
      \xi = \int_0^{\cos\theta_{\nu(\bar{\nu})}}P_n(\cos\theta_{\nu(\bar{
\nu})})
                          {\rm d}(\cos\theta_{\nu(\bar{\nu})}),
\label{eqn:c2}
    \end{equation}

\noindent where
    \begin{equation}
       P_n(\cos\theta_{\nu(\bar{\nu})}) =  I(\cos\theta_{\nu(\bar{\nu})}) 
           \Bigg/
       \int_0^1 I(\cos\theta_{\nu(\bar{\nu})}){\rm d}(\cos\theta_{\nu(\bar{
\nu})}).
\label{eqn:c3}
    \end{equation}
\noindent {\bf Procedure B}\\
\noindent
For the sampled ${\rm d}(\cos\theta_{\nu(\bar{\nu})})$ in the Procedure A, 
we sample 
$E_{\nu(\bar{\nu})}$ from Eq.(\ref{eqn:c4}) by using ${\xi}$, the uniform 
randum number between (0,1) 

 \begin{equation}
    \displaystyle
       \xi = \int_{E_{\nu(\bar{\nu}),{\rm min}}}^{E_{\nu(\bar{\nu})}} 
                    P_{pro}(E_{\nu(\bar{\nu})},\cos\theta_{\nu(\bar{\nu
})}){\rm d}E_{\nu(\bar{\nu})}, 
\label{eqn:c4}
    \end{equation}
where
      \begin{eqnarray}
\lefteqn{
         P_{pro}(E_{\nu(\bar{\nu})},t,\cos\theta_{\nu(\bar{\nu})}){\rm 
d}E_{\nu(\bar{\nu})} =
} \nonumber \\
&&  
\frac{         N_{\rm pro}( E_{\nu(\bar{\nu})},t,\cos\theta_{\nu(\bar{\nu
})}){\rm d}E_{\nu(\bar{\nu})}
}
{\displaystyle         \int_{E_{\nu(\bar{\nu}),{\rm min}}}^{E_{\nu(\bar{\nu
}),{\rm max}}} 
         N_{\rm pro}( E_{\nu(\bar{\nu})},t,\cos\theta_{\nu(\bar{\nu})}){\rm
 d}E_{\nu(\bar{\nu})}
}.
\label{eqn:c5}
      \end{eqnarray}

\noindent {\bf Procedure C}\\
\noindent 
For the sampled $E_{\nu(\bar{\nu})}$ in the Procedure B, we sample 
$E_{\mu(\bar{\mu})}$ from Eqs. (\ref{eqn:a2}) and  (\ref{eqn:a3}). For the 
sampled  $E_{\nu(\bar{\nu})}$ 
and $E_{\mu(\bar{\mu})}$, we determine $\cos{\theta}_s$, the scattering 
angle of the muon uniquely from Eq. (\ref{eqn:a1}).\\

\noindent {\bf Procedure D}\\
\noindent
We determine, $\phi$, the azimuthal angle of the scattering lepton from 
Eq. (\ref{eqn:a5}) by using ${\xi}$, an uniform randum number between 
(0,1). \\

\noindent {\bf Procedure E}\\
\noindent  
We obtain $\cos{\theta}_{\mu(\bar{\mu})}$ from Eq. (\ref{eqn:a6}).  As the 
result, we obtain a pair of ($\cos\theta_{\nu(\bar{\nu})}$, 
$\cos{\theta}_{\mu(\bar{\mu})}$) through Procedures A to E. Repeating the 
Procedures A to E, we finally the correlation between the zenith angle of 
the incident neutrino and that of the emitted muon. 

\section*{References}


\begin{thebibliography}{}
%
%
  \bibitem{Hirata}Hirata, KS {\it et al.}, Phys.Lett.{\bf B205}(1988)416\\
 Hirata, KS {\it et al.}, Phys.Lett.{\bf B280}(1992)146\\
Casper, D {\it et al.}, Phys.Rev.Lett.{\bf 66}(1991)2561\\
Becker-Szendy, R {\it et al.}, Phys.Rev. D {\bf 46}(1992)3720.
  \bibitem{Hatakeyama}Hatakeyama, S {\it et al.}, Phys.Rev.Lett.{\bf 81}(1998)2010.
  \bibitem{Kajita2}Kajita, T, Neutrino 98, Takayama, Japan, June 4-9 1998\\
Fukuda, Y, Phys.Rev.Lett.{\bf 81}(1998)1562. 
  \bibitem{Mann} Mann, WA, Nucl.Phys.Proc.Supple Vol.{\bf 91}(2000)134\\
Ambrosio, M{\it et al.}, Phys.Lett.{\bf B478}(2000)3.
  \bibitem{K2K} K2K, Phys.Rev. D{\bf 74}(2006)72003
  \bibitem{MINOS} Michael, DG {\it et al.}, Phys.Rev.Lett.{\bf 97}(2006)
  191801
  \bibitem{kajita1} Kajita, T and Totsuka, Y, Rev. Mod. Phys., {\bf 73} (2001) 85. See p. 101.
  \bibitem{ishitsuka} Ishitsuka, M, Ph.D thesis, University of Tokyo (2004). See p. 138.
  \bibitem{Jung} Jung, CK, Kajita, T, Mann, T and McGrew, C,  Anual. Rev. Mod. Sci. {\bf vol.15} (2005) 431
  \bibitem{r4} Renton, P., {\it Electro-weak Interaction}, Cambridge University Press (1990). See p. 405.
  \bibitem{honda} Honda, M., {\it et al.}, \  Phys.\ Rev. D {\bf 52} (1996) 4985
 \bibitem{Ashie} Ashie,Y. {\it et al.}, Phys. Rev. D {\bf 71} (2005) 112005.

\end{thebibliography}
\end{document}